\documentclass[amsmath,amssymb,twocolumn,prc,showpacs,nofootinbib,
floatfix,superscriptaddress]{revtex4-1}
\usepackage{graphics}
\usepackage{graphicx}
\usepackage{dcolumn}
\usepackage{bm}
\usepackage{hyperref}
\hypersetup{colorlinks,citecolor=black,filecolor=black,linkcolor=black,urlcolor=black} 
\usepackage{float}
\newcommand{\aipcp}{AIP Conf. Proc.}
\newcommand{\apparis}{Ann. Phys. (Paris)}

\newcommand{\aap}{Astron. Astrophys.}
\newcommand{\apjl}{Astrophys. J. Lett.}
\newcommand{\ieee}{IEEE Trans. Nucl. Sci.}
\newcommand{\mnras}{Mon. Not. R. Astron. Soc.}
\newcommand{\nim}{Nucl. Instrum. Methods}
\newcommand{\nima}{Nucl. Instrum. Methods A}
\newcommand{\nimb}{Nucl. Instrum. Methods B}
\newcommand{\npa}{Nucl. Phys. A}
\newcommand{\zpa}{Z. Phys. A}
\def\prc{ Phys.\ Rev.\ C }
\begin{document}
\title{Thermonuclear Reaction Rate of $^{18}$O($p$,$\gamma$)$^{19}$F}
\author{M. Q. Buckner}\email[Corresponding author:  ]{emcuebe@physics.unc.edu}
\affiliation{University of North Carolina at Chapel Hill, Chapel Hill, NC 27599-3255; and Triangle Universities Nuclear Laboratory, Durham, NC 27708-0308, USA}
\author{C. Iliadis}
\affiliation{University of North Carolina at Chapel Hill, Chapel Hill, NC 27599-3255; and Triangle Universities Nuclear Laboratory, Durham, NC 27708-0308, USA}
\author{J. M. Cesaratto}\altaffiliation{Present address:  SLAC National Accelerator Laboratory, 2575 Sand Hill Rd., MS 17,  Menlo Park, CA 94025}
\affiliation{University of North Carolina at Chapel Hill, Chapel Hill, NC 27599-3255; and Triangle Universities Nuclear Laboratory, Durham, NC 27708-0308, USA}
\author{C. Howard}
\affiliation{University of North Carolina at Chapel Hill, Chapel Hill, NC 27599-3255; and Triangle Universities Nuclear Laboratory, Durham, NC 27708-0308, USA}
\author{T. B. Clegg}
\affiliation{University of North Carolina at Chapel Hill, Chapel Hill, NC 27599-3255; and Triangle Universities Nuclear Laboratory, Durham, NC 27708-0308, USA}
\author{A. E. Champagne}
\affiliation{University of North Carolina at Chapel Hill, Chapel Hill, NC 27599-3255; and Triangle Universities Nuclear Laboratory, Durham, NC 27708-0308, USA}
\author{S. Daigle}
\affiliation{University of North Carolina at Chapel Hill, Chapel Hill, NC 27599-3255; and Triangle Universities Nuclear Laboratory, Durham, NC 27708-0308, USA}
\date{\today}
\begin{abstract}
For stars with 0.8~M$_{\odot}$~$\leq$~M~$\leq$~8.0 M$_{\odot}$, nucleosynthesis enters its final phase during the asymptotic giant branch (AGB) stage.  During this evolutionary period, grain condensation occurs in the stellar atmosphere, and the star experiences significant mass loss.  The production of presolar grains can often be attributed to this unique stellar environment.  A subset of presolar oxide grains features dramatic $^{18}$O depletion that cannot be explained by the standard AGB star burning stages and dredge-up models.  An extra mixing process, referred to as $\it{cool~bottom~processing}$ (CBP), was proposed for low-mass AGB stars.   The $^{18}$O depletion observed within certain stellar environments and within presolar grain samples may result from the $^{18}$O~+~$p$ processes during CBP.  We report here on a study of the $^{18}$O($p$,$\gamma$)$^{19}$F reaction at low energies.  Based on our new results, we found that the resonance at E$^{\mathrm{lab}}_{\mathrm{R}}$~=~95~keV has a negligible affect on the reaction rate at the temperatures associated with CBP.  We also determined that the direct capture S-factor is almost a factor of 2 lower than the previously recommended value at low energies.  An improved  thermonuclear reaction rate for $^{18}$O($p$,$\gamma$)$^{19}$F is presented.  
\end{abstract}
\pacs{}
\maketitle
\section{Introduction}\label{ss:18O}
An overwhelming majority of the matter within our solar system has a unique $^{18}$O/$^{16}$O isotopic signature.  However, a collection of presolar grain samples feature peculiar oxygen isotopic ratios.  These outliers are found within the trove of presolar grains gathered over the years from primitive meteorites and interplanetary dust particles.  This study is motivated by observations of presolar grains that nucleated in the atmospheres of distant, evolved stars before the formation of the Sun.  These grains retain the isotopic ratios of the stellar surface they originated from.  During the birth of the Sun, most presolar grains were annihilated as gas and dust collapsed to form the nascent star.  As the solar system cooled and the Sun ascended the main sequence, the presolar grains that survived were incorporated into primitive meteorites.  The study of their abnormal isotopic ratios provides crucial constraints for astrophysical models.  This paper focuses on oxide grains referred to as $\it{Group~2}$ grains, approximately 15$\%$ of all presolar oxides~\cite{NIT08}.  They exhibit a characteristic $^{18}$O/$^{16}$O abundance ratio $\leq$1.5$\times$10$^{-3}$~\cite{PAL11}, reflecting a substantial $^{18}$O depletion~\cite{HOP10} with respect to the solar value, (2.09$^{+0.13}_{-0.12}$)$\times$10$^{-3}$~\cite{SCO06}.

It has been hypothesized that asymptotic giant branch (AGB) stars are an $^{18}$O depletion site~\cite{NIT08}.  During the AGB stage---the final phase of nucleosynthesis during the evolution of a 0.8$-$8.0 M$_{\odot}$ star~\cite{LAT99,NOW01}---a star undergoes substantial nucleosynthesis and mass loss.  Peeling away the surface layers enveloping an AGB star reveals numerous burning sites and a complex interplay between these regions.  A stellar core, composed of electron degenerate carbon and oxygen, is surrounded by alternately burning helium and hydrogen shells.  During periods of helium-burning, referred to as thermal pulses, thermonuclear runaway (TNR) occurs and drives convection between the two burning sites.  When the TNR subsides, the star compensates for this period of activity by expanding and cooling.  The hydrogen burning shell is quenched during expansion, and the convective envelope dredges the products of nucleosynthesis to the surface of the star (third dredge-up).  After this dredge-up event, the star contracts, and the hydrogen shell reignites.  This interplay between the helium and hydrogen shells repeats episodically~\cite{LAT99}.  

During the AGB phase, $^{18}$O depletion may occur due to $\it{cool~bottom~processing}$ (CBP)~\cite{NIT08}.  This $\it{extra~mixing}$ process was proposed by~\citet{WAS95a} to account for isotopic anomalies, including $^{18}$O depletion, in presolar grains.  During CBP, material circulates between the convective envelope and the radiative zone that separates the envelope from the hydrogen burning shell.  The base of the convective envelope remains cool, thus distinguishing this process from $\it{hot~bottom~burning}$ that occurs in 4$-$7 M$_{\odot}$ AGB stars~\cite{WAS95a,NIT08}.  As the circulated matter approaches the hydrogen shell, it reaches temperatures high enough to destroy $^{18}$O via hydrogen burning.  The processed material is then recirculated into the convective envelope and transported to the stellar surface.  Grains nucleate in the stellar atmosphere depleted in $^{18}$O due to processes that occurred deep within the star.  Then, powerful stellar winds inject these grains into the interstellar medium. The mechanism driving CBP is not understood, and several explanations have been proposed, including magnetic buoyancy~\cite{BUS10}, gravity waves~\cite{DEN03}, shear instability~\cite{ZAH92,MAE98}, meridional circulation~\cite{SWE79}, and convective overshoot~\cite{HER97}.

The observed $^{18}$O depletion in some presolar oxide grains and AGB stellar atmospheres helped motivate the introduction of CBP into AGB stellar models.  These models provided some insight into the class of AGB stars that might experience CBP and the temperature of the stellar plasma at the site of this extra mixing.  According to~\citet{PAL11}, $^{18}$O depletion by CBP may occur in 1.5$-$1.7~M$_{\odot}$ AGB stars.  Cool bottom processing models resulted in stellar plasma temperature regimes within low-mass AGB stars that could allow for $^{18}$O~+~$p$ reactions to occur.  According to~\citet{NOL03}, a temperature in excess of 39.8$-$44.7~MK (depending on the evolution of the core, hydrogen burning shell, and convective envelope boundaries) is sufficient for $^{18}$O depletion during CBP in a 1.5~M$_{\odot}$ model star.  

The depletion of $^{18}$O in a stellar plasma at low temperatures is driven by $^{18}$O($p$,$\alpha$)$^{15}$N and, to a lesser extent, $^{18}$O($p$,$\gamma$)$^{19}$F.  The former reaction was recently studied indirectly by \citet{LAC10}.  In the present work, we report a direct, low-energy measurement of the $^{18}$O($p$,$\gamma$)$^{19}$F reaction.  The goal of this measurement was to improve our knowledge of levels in the $^{19}$F compound nucleus that are relevant to nuclear astrophysics.
\begin{figure}[!bp]
\begin{center}
\thinlines
\setlength{\unitlength}{1.1mm}
\begin{picture}(70,100)
\put(15,31.26){\line(0,1){58.73}}
\multiput(15,22)(0,4){3}{\line(0,1){2}}
\put(20,22){\line(0,1){68}}
\put(65,22){\line(0,1){68}}
\put(20,5){\line(0,1){16}}
\put(65,5){\line(0,1){16}}
\put(18,21){\line(1,0){4}}
\put(18,22){\line(1,0){4}}
\put(63,21){\line(1,0){4}}
\put(63,22){\line(1,0){4}}
\put(0,27){\large $\mathsf{^{18}O+\it{p}}$}
\put(39,0){\large $\mathsf{^{19}F}$}
\put(4.6,96){\large $\mathsf{E^{lab}_{R}}$}
\put(3.6,91){$\mathsf{[keV]}$}
\put(22,96){\large $\mathsf{E_{x}}$}
\put(20.5,91){$\mathsf{[keV]}$}
\put(58.25,96){\large $\mathsf{2J^{\pi}}$}
\put(20,5){\line(1,0){45}}
\put(22,5.25){\footnotesize $\mathsf{0}$}
\put(58.25,5.25){\footnotesize $\mathsf{1^{+}}$}
\put(20,12.33){\line(1,0){45}}
\put(22,12.58){\footnotesize $\mathsf{110}$}
\put(58.25,12.58){\footnotesize $\mathsf{1^{-}}$}
\put(20,18.13){\line(1,0){45}}
\put(22,18.38){\footnotesize $\mathsf{197}$}
\put(58.25,18.38){\footnotesize $\mathsf{5^{+}}$}
\put(20,25){\line(1,0){1.75}}
\put(22,24.25){\footnotesize $\mathsf{7900}$}
\put(27.75,25){\line(1,0){37.25}}
\put(20,26.93){\line(1,0){9.85}}
\put(35.5,26.93){\line(1,0){18}}
\put(59.75,26.93){\line(1,0){5.25}}
\put(30,25.6){{\footnotesize $\mathsf{7929}$}}
\put(53.75,25.6){\footnotesize $\mathsf{7^{+},9}$}
\put(20,27.46){\line(1,0){45}}
\put(22,27.71){\footnotesize $\mathsf{7937}$}
\put(57.1,27.71){\footnotesize $\mathsf{11^{+}}$}
\put(0,31.27){\line(1,0){15}}
\put(0,31.51){\footnotesize $\mathsf{7994}$}
\put(7,32.0){\footnotesize $\mathsf{22}$}
\put(10, 32.6){\line(1,0){0.5}}
\put(10.75, 32.6){\line(1,0){0.5}}
\put(11.5, 32.6){\line(1,0){0.5}}
\put(12.25, 32.6){\line(1,0){0.5}}
\put(13, 32.6){\line(1,0){0.5}}
\put(15, 32.6){\vector(1, 0){0}}
\put(20,32.6){\line(1,0){45}}
\put(22,32.85){\footnotesize $\mathsf{8014}$}
\put(58.25,32.85){\footnotesize $\mathsf{5^{+}}$}
\put(7,36.66){\footnotesize $\mathsf{95}$}
\put(10, 37.26){\line(1,0){0.5}}
\put(10.75, 37.26){\line(1,0){0.5}}
\put(11.5, 37.26){\line(1,0){0.5}}
\put(12.25, 37.26){\line(1,0){0.5}}
\put(13, 37.26){\line(1,0){0.5}}
\put(15, 37.26){\vector(1, 0){0}}
\put(20,37.26){\line(1,0){45}}
\put(22,37.51){\footnotesize $\mathsf{8084}$}
\put(58.25,37.51){\footnotesize $\mathsf{3^{+}}$}
\put(5.8,40.26){\footnotesize $\mathsf{151}$}
\put(10, 40.86){\line(1,0){4}}
\put(15, 40.86){\vector(1, 0){0}}
\put(20,40.86){\line(1,0){1.75}}
\put(22,40.11){\footnotesize $\mathsf{8138}$}
\put(58.25,40.11){\footnotesize $\mathsf{1^{+}}$}
\put(27.75,40.86){\line(1,0){30.25}}
\put(61,40.86){\line(1,0){4}}
\put(20,42.33){\line(1,0){45}}
\put(22,42.58){\footnotesize $\mathsf{8160}$}
\put(20,44.93){\line(1,0){45}}
\put(22,45.18){\footnotesize $\mathsf{8199}$}
\put(57.3,45.18){\footnotesize $\mathsf{(5^{+})}$}
\put(5.8,44.33){\footnotesize $\mathsf{217}$}
\put(10, 44.93){\line(1,0){4}}
\put(15, 44.93){\vector(1, 0){0}}
\put(20,44.93){\line(1,0){1.75}}
\put(5.8,48){\footnotesize $\mathsf{275}$}
\put(10, 48.6){\line(1,0){4}}
\put(15, 48.6){\vector(1, 0){0}}
\put(20,48.6){\line(1,0){1.75}}
\put(22,48.1){\footnotesize $\mathsf{8254}$}
\put(53.7,48.1){\footnotesize $\mathsf{(5,7)^{-}}$}
\put(27.75,48.6){\line(1,0){25.5}}
\put(61,48.6){\line(1,0){4}}
\put(20,50.86){\line(1,0){1.75}}
\put(22,50.25){\footnotesize $\mathsf{8288}$}
\put(27.75,50.86){\line(1,0){29}}
\put(61,50.86){\line(1,0){4}}
\put(56.9,50.25){\footnotesize $\mathsf{13^{-}}$}
\put(5.8,51.73){\footnotesize $\mathsf{334}$}
\put(10, 52.33){\line(1,0){4}}
\put(15, 52.33){\vector(1, 0){0}}
\put(20,52.33){\line(1,0){45}}
\put(22,52.58){\footnotesize $\mathsf{8310}$}
\put(58.25,52.58){\footnotesize $\mathsf{5^{+}}$}
\put(20,56.33){\line(1,0){45}}
\put(22,56.58){\footnotesize $\mathsf{8370}$}
\put(53.75,56.58){\footnotesize $\mathsf{5^{+},7}$}
\put(5.8,69){\footnotesize $\mathsf{623}$}
\put(10, 70.6){\line(1,0){4}}
\put(15, 70.6){\vector(1, 0){0}}
\put(20,70.6){\line(1,0){45}}
\put(22,68.5){\footnotesize $\mathsf{8584}$}
\put(58.25,68.5){\footnotesize $\mathsf{5^{+}}$}
\put(5.8,71){\footnotesize $\mathsf{632}$}
\put(10, 71.13){\line(1,0){4}}
\put(15, 71.13){\vector(1, 0){0}}
\put(20,71.13){\line(1,0){1.75}}
\put(22,71){\footnotesize $\mathsf{8592}$}
\put(58.25,71){\footnotesize $\mathsf{3^{-}}$}
\put(27.75,71.13){\line(1,0){30.25}}
\put(61,71.13){\line(1,0){4}}
\put(20,73.6){\line(1,0){1.75}}
\put(22,73){\footnotesize $\mathsf{8629}$}
\put(27.75,73.6){\line(1,0){30.25}}
\put(61,73.6){\line(1,0){4}}
\put(58.25,73){\footnotesize $\mathsf{7^{-}}$}
\put(5.8,74.4){\footnotesize $\mathsf{693}$}
\put(10, 75){\line(1,0){4}}
\put(15, 75){\vector(1, 0){0}}
\put(20,75){\line(1,0){45}}
\put(22,75.25){\footnotesize $\mathsf{8650}$}
\put(58.25,75.25){\footnotesize $\mathsf{1^{+}}$}
\put(5.8,83.93){\footnotesize $\mathsf{844}$}
\put(10, 84.53){\line(1,0){4}}
\put(15, 84.53){\vector(1, 0){0}}
\put(20,84.53){\line(1,0){45}}
\put(22,84.78){\footnotesize $\mathsf{8793}$}
\put(58.25,84.78){\footnotesize $\mathsf{1^{+}}$}
\end{picture}
\caption{\label{fig:lev}Truncated $^{19}$F level diagram and $^{18}$O~+~$p$ resonances~\cite{SYM78,LOR79,WIE80,VOG90,TIL95,BEC95} through E$^{\mathrm{lab}}_{\mathrm{R}}$~=~844~keV.  The E$^{\mathrm{lab}}_{\mathrm{R}}$~=~95~keV resonance corresponds to the E$_{\mathrm{x}}$~=~8084~keV excited state.  For an explanation of our reported spin and parity for this level, see Sec.~\ref{ss:18O}.  Dashed arrows indicate unobserved resonances and solid arrows indicate observed resonances with known $\gamma$-ray decays.  The proton threshold, Q$_{p\gamma}$~=~7994~keV, was taken from Ref.~\cite{AUD11}.}
\end{center}
\end{figure}
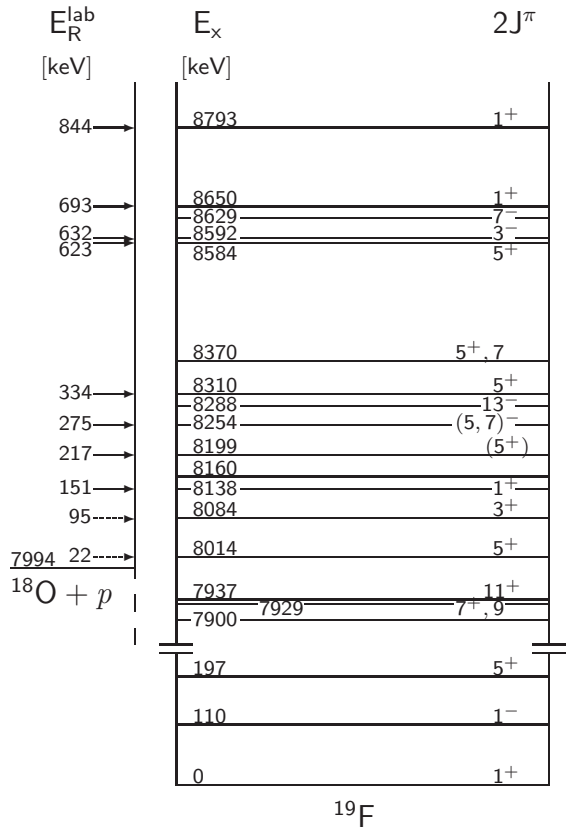    

Within the CBP temperature regime, the $^{18}$O($p$,$\gamma$)$^{19}$F reaction rate may be influenced by an unobserved, low-energy resonance at E$^{\mathrm{lab}}_{\mathrm{R}}$~=~95~$\pm$~3~keV~\cite{TIL95,AUD11} (see Fig.~\ref{fig:lev}).  In the competing $^{18}$O($p$,$\alpha$)$^{15}$N reaction, a strength of $\omega\gamma_{p\alpha}$~=~(1.6~$\pm$~0.5)$\times$10$^{-7}$~eV at E$^{\mathrm{lab}}_{\mathrm{R}}$~=~95~keV was directly measured  by~\citet{LOR79}.  The resonance strength is the integral of the reaction cross section and is defined as~\cite{ILI07}:
\begin{equation}\label{wgpx}
\omega\gamma~=~\frac{(2\mathrm{J}~+~1)}{(2\mathrm{J}_{p}~+~1)(2\mathrm{J}_{\mathrm{t}}~+~1)}\frac{\Gamma_{p}\Gamma_{\mathrm{x}}}{(\Gamma_{p}~+~\Gamma_{\alpha}~+~\Gamma_{\gamma})}
\end{equation}
where J is the compound nucleus spin, J$_{p}$~=~1/2 is the proton spin, J$_{\mathrm{t}}$~=~0 is the $^{18}$O target nucleus spin, $\Gamma_{p}$ is the proton partial width, $\Gamma_{\gamma}$ is the $\gamma$-ray partial width, $\Gamma_{\alpha}$ is the $\alpha$-particle partial width, $\Gamma_{\mathrm{x}}~=~\Gamma_{\gamma}$ for photon emission, and $\Gamma_{\mathrm{x}}~=~\Gamma_{\alpha}$ for $\alpha$-particle emission.  In the $^{18}$O($p$,$\gamma$)$^{19}$F reaction, the E$^{\mathrm{lab}}_{\mathrm{R}}$~=~95~keV resonance has never been observed, and none of the $\gamma$-ray decays from the resonance level are known.  Upper limits were placed on the resonance strength in the past, first by~\citet{WIE80} with $\omega\gamma_{p\gamma}$~$\leq$~5$\times$10$^{-8}$~eV and then by~\citet{VOG90} with $\omega\gamma_{p\gamma}$~$\leq$~4$\times$10$^{-8}$~eV.  With a proton separation energy of Q$_{p\gamma}$~=~7993.5994~$\pm$~0.0011~keV~\cite{AUD11}, the E$^{\mathrm{lab}}_{\mathrm{R}}$~=~95~keV resonance corresponds to the E$_{\mathrm{x}}$~=~8084~$\pm$~3~keV~\cite{TIL95} level in the $^{19}$F nucleus.  

The previous experimental information regarding the structure of this compound nucleus level is summarized in Table~\ref{table:levparam}.  From the  $^{18}$O($^{3}$He,$d$)$^{19}$F experiment performed by~\citet{SCH70}, it is clear that the proton angular momentum transfer for the 8084~keV level is restricted to ${\ell_{p}}$~=~(2,~3).  In~\citet{LAC08}, the Trojan Horse Method was used to investigate this level with the $^{2}$H($^{18}$O,$\alpha$$^{15}$N)n reaction.  They determined that J~=~3/2 and $\ell_{\alpha}$~=~1.  Consequently, based on the angular momentum coupling rules, the spin-parity and orbital angular momentum amount to J$^{\pi}$~=~(3/2)$^{+}$ and $\ell_{p}$~=~2, respectively.  Note that an incorrect spin and parity of J$^{\pi}$~=~(3/2)$^{-}$ was assumed previously for this level~\cite{ILI10b}.  
\begin{table}[!bp]
\begin{center}
\caption{\label{table:levparam}E$_{\mathrm{x}}$~=~8084~keV level parameters.}
\begin{tabular}{cccccc} 
\hline\hline
 \multicolumn{2}{c}{Parameter} & \multicolumn{2}{c}{Value (eV)} & \multicolumn{2}{c}{Reference} \\
\hline
  \multicolumn{2}{c}{$\omega\gamma_{p\alpha}$} & \multicolumn{2}{c}{(1.6~$\pm$~0.5)$\times$10$^{-7}$} & \multicolumn{2}{c}{~\cite{LOR79}} \\
 \multicolumn{2}{c}{$\omega\gamma_{p\gamma,\mathrm{UL}}$} & \multicolumn{2}{c}{$\leq$ 4.0$\times$10$^{-8}$} & \multicolumn{2}{c}{~\cite{VOG90}} \\
 \multicolumn{2}{c}{$\omega\gamma_{p\gamma,\mathrm{LL}}$} & \multicolumn{2}{c}{$\geq$ 1.3$\times$10$^{-11}$} & \multicolumn{2}{c}{Sec.~\ref{ss:rr}} \\
  \multicolumn{2}{c}{$\Gamma_{\gamma}$} & \multicolumn{2}{c}{(6.0~$\pm$~2.5)$\times$10$^{-1}$~\footnote{Private communication from K. Allen quoted in~\citet{WIE80}.}} & \multicolumn{2}{c}{~\cite{WIE80}} \\
 \multicolumn{2}{c}{$\Gamma$} & \multicolumn{2}{c}{$\leq$ 3.0$\times$10$^{3}$~\footnote{Total width determined from slope of front edge of thick-target yield curve.}} & \multicolumn{2}{c}{~\cite{LOR79}} \\
\hline\hline
\end{tabular}
\end{center}
\end{table}

Here we report on a new search for the E$^{\mathrm{lab}}_{\mathrm{R}}$ = 95 keV resonance in $^{18}$O($p$,$\gamma$)$^{19}$F with significantly improved sensitivity compared to previous studies \cite{WIE80,VOG90}.  In the following we will discuss the experimental setup, including a brief outline of the accelerators (Sec.~\ref{ss:lena}) and detector system (Sec.~\ref{ss:detect}).  Our oxygen target fabrication process is described in Sec.~\ref{ss:trgt}.  The methodology we employed to characterize our detector efficiencies is discussed in Sec.~\ref{ss:detect}, and Sec.~\ref{ss:daq} outlines some of the features of our data acquisition electronics.  Results for resonant and non-resonant proton capture on $^{18}$O are presented in Sec.~\ref{ss:res} and Sec.~\ref{ss:nonres}, respectively.  An improved $^{18}$O($p$,$\gamma$)$^{19}$F reaction rate is presented in Sec.~\ref{ss:rr}.  Concluding remarks are presented in Sec.~\ref{ss:conc}.    
\section{Experiment}
\subsection{Accelerators}\label{ss:lena}
The Laboratory for Experimental Nuclear Astrophysics (LENA) is dedicated to the measurement of low-energy nuclear reactions relevant to stellar nucleosynthesis.  The cross sections measured at LENA lie within an energy regime that is susceptible to Coulomb suppression, and the LENA facility features key tools that increase the detection sensitivity.  

The LENA facility is a two-accelerator laboratory and consists of a high-current, low-energy Electron Cyclotron Resonance Ion Source (ECRIS) and an upgraded HVEC 1MV JN Van de Graaff.  The LENA ECRIS produces average beam currents of I$_{p}$~=~1.5~mA on target within a bombarding energy range of 50~keV~$\leq$~E$^{\mathrm{lab}}_{p}$~$\leq$~215~keV.  The high-current allows for a substantial increase in low-energy nuclear reaction yields.  The LENA 1MV JN Van de Graaff is capable of producing H$^{+}$ beam currents of  I$_{p}$~$\leq$~250 $\mu$A at the target.  Typical beam energy resolution achieved with the JN ranges between 1$-$2~keV.  In this study, it was primarily used to test our targets by measuring excitation functions ($\gamma$-ray yield vs. bombarding energy) of the well-known $^{18}$O($p$,$\gamma$)$^{19}$F resonance at E$^{\mathrm{lab}}_{\mathrm{R}}$~=~150.82~$\pm$~0.09~keV~\cite{BEC95}.  These excitation functions provided information on target thickness and stability during the experiment.  See Sec.~\ref{ss:trgt} for more information on our $^{18}$O targets.  A detailed description of the LENA accelerator facility can be found in~\citet{CES10}.
\subsection{Targets}\label{ss:trgt}
The anodic oxidation of tantalum targets was first outlined by~\citet{AMS64}.  The anodization process allows target thicknesses to be consistently reproduced, and it also allows the production of robust oxygen targets that remain stable when exposed to intense H$^{+}$ beam.  A new anodization chamber was designed and assembled for our measurement according to the description in Ref.~\cite{AMS78b}.     

During fabrication, all tantalum backings were etched in an acid bath in order to reduce beam induced backgrounds by removing surface contaminants ($^{11}$B and $^{19}$F).  Subsequently, all etched tantalum backings were resistively heated.  These outgassed target backings were anodized at 64~V using 99.3$\%$ enriched $^{18}$O water to produce Ta$_{2}$$^{18}$O$_{5}$ targets with an expected target thickness of $\sim$18~keV at E$^{\mathrm{lab}}_{\mathrm{R}}$~=~151~keV.  

Excitation functions were collected during this experiment at the well-known E$^{\mathrm{lab}}_{\mathrm{R}}$~=~151~keV resonance in the $^{18}$O($p$,$\gamma$)$^{19}$F reaction with the JN Van de Graaff.  Target thicknesses near 100~keV were estimated with the relationship~\cite{ILI07}: 
\begin{equation}\label{trgtthcknss}
\frac{\Delta \mathrm{E}(151)}{\epsilon_{\mathrm{eff}}(151)}~=~\frac{\Delta \mathrm{E}(\mathrm{E}_{p})}{\epsilon_{\mathrm{eff}}(\mathrm{E}_{p})}
\end{equation}
where $\Delta$E is the measured target thickness in energy units, and $\epsilon_{\mathrm{eff}}$ is the effective stopping power in the center-of-mass system, derived from Bragg's rule~\cite{FOX05,ILI07}:
\begin{equation}\label{eps_eff}
\epsilon_{\mathrm{eff}}~=~\frac{\mathrm{M}_{\mathrm{^{18}O}}}{\mathrm{M}_{\mathit{p}}+\mathrm{M}_{\mathrm{^{18}O}}}\Big(\frac{\mathrm{N}_{\mathrm{O}}}{\mathrm{N}_{\mathrm{^{18}O}}}\epsilon_{\mathrm{^{18}O}}~+~\frac{\mathrm{N}_{\mathrm{Ta}}}{\mathrm{N}_{\mathrm{^{18}O}}}\epsilon_{\mathrm{Ta}}\Big)
\end{equation}
where M$_{p}$ and M$_{\mathrm{^{18}O}}$ are the mass of the proton and the $^{18}$O atom, $\epsilon_{\mathrm{^{18}O}}$ and $\epsilon_{\mathrm{Ta}}$ are the laboratory stopping powers of protons in $^{18}$O and Ta (calculated with $\mathtt{SRIM}$~\cite{ZIE04}), and N$_{i}$ are number densities (N$_{\mathrm{O}}$~=~N$_{\mathrm{^{16}O}}$~+~N$_{\mathrm{^{17}O}}$~+~N$_{\mathrm{^{18}O}}$).  We found that our targets could withstand proton accumulations of Q$_{p}$ $>$ 45 C without significant degradation at I$^{\mathrm{ECRIS}}_{p}$~=~0.5$-$1.0~mA.  
\subsection{Detectors}\label{ss:detect}
Almost all $^{18}$O($p$,$\gamma$)$^{19}$F resonances are known to decay via emission of multiple, coincident $\gamma$-rays.  Therefore, the simultaneous detection of two or more photons allows an opportunity to increase the signal-to-noise ratio significantly.  To accomplish this signal optimization, a $\gamma$-ray spectrometer consisting of several detectors was used.  The LENA $\gamma\gamma$-coincidence detector system was assembled with an 135$\%$ HPGe detector at 0 degrees to the beam and in close running geometry with the target chamber.  The distance between the HPGe detector and the target midpoint was 1.1 cm \cite{LON06}.  The target chamber and HPGe detector were surrounded by a 16-segment NaI(Tl) annulus.  Plastic scintillator paddles covered the two detectors on five sides and suppressed cosmic-ray muon events.  In a two-dimensional NaI(Tl) vs. HPGe energy spectrum, appropriate gates were set during off-line data sorting.  The low-energy thresholds set on these gates removed events caused by environmental background ($^{40}$K, $^{208}$Tl), and the high-energy thresholds excluded events with a total energy that exceeded the excitation energy of the decaying $^{19}$F compound nucleus.  Most of the latter events were presumably caused by cosmic ray interactions.  The LENA $\gamma\gamma$-coincidence detector was described in detail by~\citet{LON06}.  

The internal geometry of the HPGe detector is well known and was measured previously~\cite{CAR10} using computed tomography (CT).  Based on the known dimensions, relative peak efficiencies were simulated using G$\textsc{eant}$4~\cite{AGO03,ALL06} by assuming mono-energetic $\gamma$-rays (E$_{\gamma}$~=~0.05$-$15.0 MeV) emitted from an extended beamspot (1.2 cm diameter) on the target.  The sum-peak method~\cite{KIM03,ILI07} was used to obtain absolute peak and total efficiencies for $^{60}$Co (E$_{\gamma}$~=~1173~keV, 1332~keV).  The absolute peak efficiency of the LENA HPGe detector was determined to be $\eta^{\mathrm{Ge,P}}_{1332}$~=~0.040~$\pm$~0.003, and the absolute total efficiency was $\eta^{\mathrm{Ge,T}}_{1253}$~=~0.188~$\pm$~0.012.  Corrections for the finite beamspot size were estimated with G$\textsc{eant}$4~\cite{LON06}. 

Peak efficiencies at lower energies were measured with $^{56}$Co and at higher energies (up to 10 MeV) with nuclear reactions.  These reactions included $^{14}$N($p$,$\gamma$)$^{15}$O at E$^{\mathrm{lab}}_{\mathrm{R}}$~=~278~keV~\cite{MAR11} and $^{27}$Al($p$,$\gamma$)$^{28}$Si at E$^{\mathrm{lab}}_{\mathrm{R}}$~=~406~keV~\cite{END90,POW98}.  All efficiency data were corrected for coincidence summing effects using the matrix method  outlined by~\citet{SEM90}.  The separate data sets were bootstrapped together across the full energy range to determine the experimental peak efficiencies of the detector.  Between measured energies, the peak efficiencies were found by interpolation using G$\textsc{eant}$4.  Total efficiencies were also simulated in G$\textsc{eant}$4 and then normalized to the measured $^{60}$Co sum-peak total efficiency.  The estimation of the $\gamma$-ray coincidence efficiencies needed for the data analysis is detailed in Sec.~\ref{ss:res}.     
\subsection{Data Acquisition and Procedure}\label{ss:daq}
As discussed in Sec.~\ref{ss:detect}, the LENA spectrometer is composed of three different detectors: a HPGe detector, a 16-segment NaI(Tl) annulus, and plastic scintillator plates.  Timing and energy signals were processed using standard NIM and VME modules.  The HPGe signals served as master triggers for the electronics.  Coincidence and anti-coincidence events were sorted using the acquisition software $\mathtt{JAM}$~\cite{SWA02}; this system was convenient for applying the desired software energy and timing gates.  Additional details concerning the electronics setup for the LENA $\gamma\gamma$-coincidence detector can be found in~\citet{LON06}.  

Initial excitation functions were produced with the JN Van de Graaff at the well-known E$^{\mathrm{lab}}_{\mathrm{R}}$~=~151~keV resonance for each target.  In order to search for the E$^{\mathrm{lab}}_{\mathrm{R}}$~=~95~keV resonance, a charge of 80~C was accumulated on-resonance at a bombarding energy of E$^{\mathrm{lab}}_{p}$~=~105~keV, and 40~C were accumulated off-resonance at E$^{\mathrm{lab}}_{p}$~=~85~keV.  The average beam current on target amounted to I$_{p}$~=~754~$\mu$A.  
\section{Results}
\subsection{Resonant Data Analysis}\label{ss:res}
\begin{figure}[!bp]
\begin{center}
\includegraphics[scale=0.45]{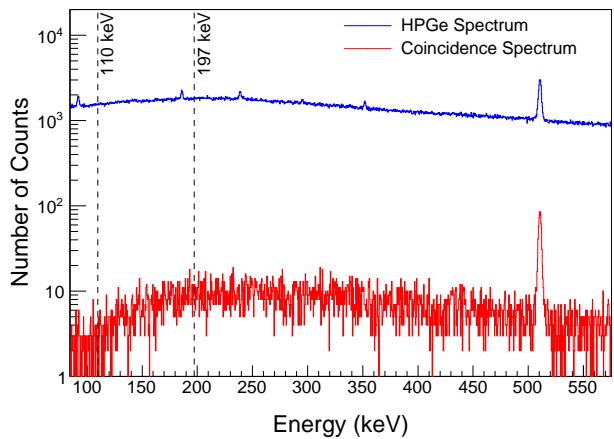}
\caption{\label{fig:coin}(Color online) HPGe singles spectrum (blue) and $\gamma\gamma$-coincidence spectrum (red).  Reduction in background amounts to a factor of 100.  The prominent background peak at 511~keV arises from the annihilation of pair-produced positrons.  Dashed lines indicate the anticipated locations of the 1~$\rightarrow$~0 (110~keV) and 2~$\rightarrow$~0 (197~keV) transitions in $^{19}$F.  The spectra shown represent on-resonance data, with a total charge accumulation of 80 C at E$^{\mathrm{lab}}_{p}$ = 105~keV.}
\end{center}
\end{figure} 
Gates were constructed in $\mathtt{JAM}$ to produce $\gamma\gamma$-coincidence spectra, uncover the $\gamma$-ray decay fingerprint of the resonance, and reduce background contributions.  Sample spectra are displayed in Fig.~\ref{fig:coin}, showing the on-resonance (ungated) singles HPGe detector spectrum in blue and the coincidence gated spectrum in red.  For the latter spectrum, only those events in the HPGe detector are accepted that are coincident with events in the NaI(Tl) counter of energy 4.25~MeV~$\leq$~E$^{\mathrm{NaI(Tl)}}_{\gamma}$~$\leq$~10.0~MeV.  It can be seen in Fig.~\ref{fig:coin} that this condition suppresses the environmental background by two orders of magnitude.  Most $^{19}$F levels decay by $\gamma\gamma$-cascades through the first (110~keV) excited state, and all $^{19}$F levels (with known decay schemes) de-excite through the second (197~keV) excited state.  In Fig.~\ref{fig:coin}, vertical dashed lines indicate anticipated locations of the $\gamma$-rays originating from the de-excitation of the first excited state (110~keV) and second excited state (197~keV).  Note that because of their low energy, the 110~keV photons would be significantly attenuated.  No peaks were observed for these two secondary decays.  In fact, although we  achieved a considerably improved detection sensitivity compared to previous studies (by about half an order of magnitude; see below), no $\gamma$-rays from the decay of the E$^{\mathrm{lab}}_{\mathrm{R}}$ = 95 keV resonance were observed in any of our singles or coincidence spectra.   

An improved upper limit on the resonance strength of the unobserved E$^{\mathrm{lab}}_{\mathrm{R}}$~=~95~keV resonance was determined relative to the strength of the well-known resonance at E$^{\mathrm{lab}}_{\mathrm{R}}$~=~151~keV.  The resonance strength is given by~\cite{GOV59,ILI07}:
\begin{equation}
\label{eqn:resstreng}
\omega\gamma~=~\frac{2 \epsilon_{\mathrm{eff}}}{\lambda^{2}}\frac{\mathcal{N}_{\mathrm{max}}}{\mathcal{N}_{p} \mathcal{B} \eta \mathcal{W}}
\end{equation}
where $\epsilon_{\mathrm{eff}}$ is the effective stopping power at the resonance energy as defined in Eq.~(\ref{eps_eff}), $\lambda$ is the de Broglie wavelength, where~\cite{ILI07}:
\begin{equation}
\label{eqn:debroglie}
\frac{\lambda^{2}}{2}~=~\Bigg(\frac{\mathrm{M}_{p}+\mathrm{M}_{\mathrm{t}}}{\mathrm{M}_{p} \mathrm{M}_{\mathrm{t}}}\Bigg) \frac{4.125\times 10^{-18}}{\mathrm{E}_{\mathrm{R}}^{\mathrm{c.m.}}}~(\mathrm{cm^{2}}),
\end{equation}
$\mathcal{N}_{\mathrm{max}}$ is the total number of detected $\gamma$-rays if the target is considered infinitely thick, $\mathcal{N}_{p}$ is the number of incident protons:
\begin{equation}
\label{eqn:particles}
\mathcal{N}_{p}~=~\frac{\mathrm{Q}}{e}
\end{equation}
where Q is the accumulated charge on target and $e$ is the unit charge in Coulomb, $\mathcal{B}$ is the branching ratio, $\eta$ is the efficiency of the detector, and $\mathcal{W}$ is the angular correlation.  For the ratio of resonance strengths we obtain~\cite{ILI07}:  
\begin{equation}
\label{eqn:relres}
\frac{\omega\gamma_{95}}{\omega\gamma_{151}}~=~ \Bigg(\frac{\epsilon_{\mathrm{eff}}\mathcal{N}_{\mathrm{max}}}{\lambda^{2}\mathcal{N}_{p}\mathcal{B}\eta\mathcal{W}}\Bigg)_{95}\times\Bigg(\frac{\epsilon_{\mathrm{eff}}\mathcal{N}_{\mathrm{max}}}{\lambda^{2}\mathcal{N}_{p}\mathcal{B}\eta\mathcal{W}}\Bigg)_{151}^{-1}.  
\end{equation}
In this equation, $\omega\gamma_{151}$~=~(9.7~$\pm$~0.5)$\times$10$^{-4}$~eV~\cite{ILI07} from the weighted mean of the resonance strengths reported in Refs.~\cite{WIE80,BEC82,VOG90}.  All $^{19}$F levels decay through the second excited state (2~$\rightarrow$~0), and we chose not to exclude the possibility that the 8084 keV level decays with a substantial primary ground state branch.  Therefore, we used the following expression to estimate an upper limit for the number of $^{19}$F compound nuclei produced~\cite{ROW02,ILI07}: 
\begin{equation}
\label{eqn:ulresfrac}
\Bigg(\frac{\mathcal{N}_{\mathrm{max}}}{\mathcal{B}\eta\mathcal{W}}\Bigg)_{95}~=~\frac{\mathcal{N}_{\mathrm{R}0}}{\eta_{\mathrm{R}0}^{\mathrm{\mathrm{Ge,P}}}}~+~\frac{\mathcal{N}_{20}}{\eta_{20}^{\mathrm{Ge,P}}f_{\gamma}}
\end{equation}
where $\mathcal{N}_{\mathrm{R}0}$ is the upper limit on the intensity of the ground state transition in the singles HPGe spectrum, $\mathcal{N}_{20}$ is the upper limit on the intensity of the decay from the $^{19}$F second excited state to the ground state (2~$\rightarrow$~0; see Fig.~\ref{fig:lev}) in the coincidence-gated HPGe spectrum, $\eta_{\mathrm{R}0}^{\mathrm{Ge,P}}$ is the HPGe peak efficiency for the ground state transition, $\eta_{20}^{\mathrm{Ge,P}}$ is the HPGe peak efficiency of the 2~$\rightarrow$~0 transition, and $f_{\gamma}$ is a $\gamma\gamma$-coincidence correction factor that depends on the $\gamma$-ray decay scheme and the coincidence gate selected.  

To calculate the correction factor $f_\gamma$, a G$\textsc{eant}$4 simulation was conducted that, for a given energy level, used the known emission probabilities to predict the total number of detected $\gamma$-rays arising from the 2~$\rightarrow$~0 transition for a variety of coincidence gates.  Our new upper limit was extracted by requiring a rectangular energy gate of 4.25~MeV~$\leq$~E$^{\mathrm{NaI(Tl)}}_{\gamma}$~$\leq$~10.0~MeV in the two-dimensional NaI(Tl) vs. HPGe coincidence energy spectrum.  The simulated coincidence histograms could then be sorted with the same energy gates and conditions that were used to analyze the experimental data.  The correction factor, $f_\gamma$, was calculated by solving the following equation~\cite{ILI07}:
\begin{equation}
\label{eqn:postpro}
\mathcal{N}'_{20}~=~\mathcal{N}_{\mathrm{R}}\eta^{\mathrm{Ge,P}}_{20}\mathit{f}_{\gamma}
\end{equation}
where $\mathcal{N}'_{20}$ is the simulated intensity of the 197~keV peak in the coincidence spectrum, $\mathcal{N}_{\mathrm{R}}$ is the total number of simulated reactions, and $\eta^{\mathrm{Ge,P}}_{20}$ is the 2~$\rightarrow$~0 singles peak efficiency (Sec.~\ref{ss:detect}).  This procedure was tested at the E$^{\mathrm{lab}}_{\mathrm{R}}$~=~151~keV resonance, where the simulated intensities agreed with the experimental values within uncertainty (4$\%$ for the 2$\rightarrow$0 decay in a rectangular coincidence spectrum).  

\begin{figure}[!bp]
\begin{center}
\includegraphics[scale=0.45]{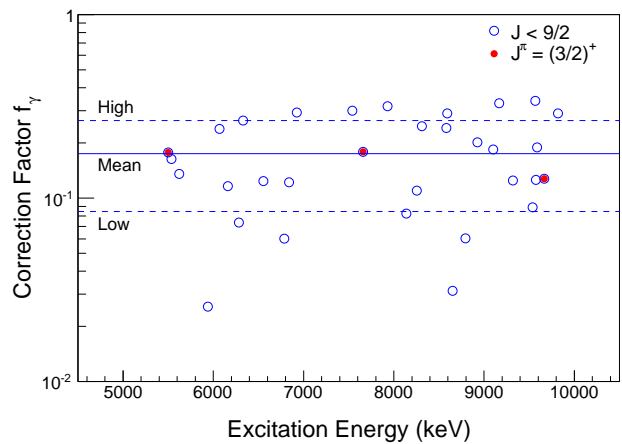}
\caption{\label{fig:fg}{(Color online) $\gamma\gamma$-coincidence correction factors ($f_{\gamma}$) for all $^{19}$F energy levels with J~$< $~9/2 (open blue circles) and E$_{\mathrm{x}}$~$\geq$~5500~keV; levels with high ground state decay modes were excluded.  Additionally, correction factors for levels with J$^{\pi}$~=~(3/2)$^{+}$ are indicated by solid red circles.  These correction factors were generated by applying a 4.25~MeV~$\leq$~E$^{\mathrm{NaI(Tl)}}_{\gamma}$~$\leq$~10.0 MeV gate.  The mean $f_{\gamma}$ value for the entire distribution, $f_{\gamma}$~=~0.17~$\pm$~0.09, is represented in this figure by the solid blue line.  The two dashed blue lines represent the uncertainty (in this instance, the root-mean-square).}} 
\end{center}
\end{figure} 
Because the E$^{\mathrm{lab}}_{\mathrm{R}}$~=~95~keV resonance has no known decay scheme or branching ratios, we first calculated values of $f_\gamma$, according to the procedure described above, for all bound and unbound $^{19}$F levels with known decay schemes~\cite{TIL95}.  We then adopted a reasonable average from the ensemble of values.  The statistical analysis was restricted to levels with J~$<$~9/2 (open blue circles in Fig.~\ref{fig:fg}) and E$_{\mathrm{x}}$~$\geq$~5500~keV.  The constraint on the spin was chosen to associate the calculated mean $f_{\gamma}$ value with low-spin states, while the energy threshold was set so that the $f_{\gamma}$ values were associated with complex $\gamma$-ray decay routes to the $^{19}$F second excited state.  As an additional constraint, no level with a ground state branching ratio that exceeded the total probability of emission to the $^{19}$F second excited state, was included in this analysis.  This final constraint was added because the ground state decay mode is already included in the strength upper limit calculation---see Eq.~(\ref{eqn:ulresfrac}).  Results of this analysis for a 4.25~MeV~$\leq$~E$^{\mathrm{NaI(Tl)}}_{\gamma}$~$\leq$~10.0 MeV gate are shown in Fig.~\ref{fig:fg}.  An average value of $f_{\gamma}$~=~0.17~$\pm$~0.09 represents a reasonable $\gamma\gamma$-coincidence correction factor estimate for the E$^{\mathrm{lab}}_{\mathrm{R}}$~=~95~keV resonance.  The quoted uncertainty of the mean correction factor is the root-mean-square of the distribution.  In Fig.~\ref{fig:fg}, the J$^{\pi}$~=~(3/2)$^{+}$ levels (levels with the same spin and parity as the E$^{\mathrm{lab}}_{\mathrm{R}}$~=~95~keV resonance level) are indicated with closed red circles.

The peak intensity upper limit for the 2$\rightarrow$0 transition (197~keV) was obtained from the HPGe coincidence spectrum using the Bayesian statistical approach outlined in~\citet{ZHU07}.  According to this method, conditional, non-informative posterior probability density functions were generated for each energy region, and peak intensity upper limits were calculated.  The E$^{\mathrm{lab}}_{\mathrm{R}}$~=~95~keV resonance strength upper limit was then determined by generating normally distributed probability density functions for all of the other quantities that entered into the resonance strength calculation---Eqs.~(\ref{eqn:debroglie}$-$\ref{eqn:ulresfrac}).  All probability density functions were then randomly sampled iteratively, and this process produced a resonance strength probability density function that was then integrated to the 90$\%$ confidence level.  A new resonance strength upper limit of $\omega\gamma_{95}$~$\leq$~7.8$\times$10$^{-9}$~eV (90$\%$~CL) was obtained for the E$^{\mathrm{lab}}_{\mathrm{R}}$~=~95~keV resonance in the $^{18}$O($p$,$\gamma$)$^{19}$F reaction.  This new upper limit improves upon the upper limit presented in~\citet{VOG90} by about a factor of 5. 
 
\subsection{Non-resonant Data Analysis}\label{ss:nonres}
Data collected during this study at E$^{\mathrm{lab}}_{p}$~=~105~keV are also important for obtaining improved estimates for the direct capture cross section of $^{18}$O($p$,$\gamma$)$^{19}$F.  During direct capture, a final bound state is created when the nucleus acquires a proton via $\gamma$-ray emission; this process occurs without the formation of a compound nucleus~\cite{ILI04}.  The experimental $^{18}$O($p$,$\gamma$)$^{19}$F direct capture cross section at E$^{\mathrm{lab}}_{p}$~=~1850~keV was measured previously by~\citet{WIE80}.  

While the $f_{\gamma}$ estimation explained in Sec.~\ref{ss:res} relied upon a statistical argument, no such assumption was necessary to determine the direct capture correction factor, $f_{\gamma}^{\mathrm{DC}}$.  To calculate the required direct capture branching ratios, we first extrapolated the experimental cross section to E$^{\mathrm{lab}}_{p}$~=~105~keV for all direct capture transitions observed in Ref.~\cite{WIE80}.  To this end, two different direct capture codes were employed.  The code $\mathtt{TEDCA}$~\cite{KRA92} was used to compute the direct capture cross section for a zero scattering potential.  The bound state and scattering state potential parameters used were adopted from~\citet{ILI04}.  The code $\mathtt{DIRCAP}$~\cite{ILI04} was utilized to perform the same calculation with a hard-sphere scattering potential.  The calculated cross sections (from E$^{\mathrm{c.m.}}_{p}$~=~0.03$-$1.99~MeV) were normalized to the measured direct capture cross sections at E$^{\mathrm{lab}}_{p}$~=~1850~keV~\cite{WIE80}.  The direct capture branching ratios derived from this procedure were required for the calculation of coincidence efficiency correction factors, $f_{\gamma}^{\mathrm{DC}}$, using the code G$\textsc{eant}$4 (Sec.~\ref{ss:res}).  

\begin{figure}[!bp]
\begin{center}
\includegraphics[scale=0.45]{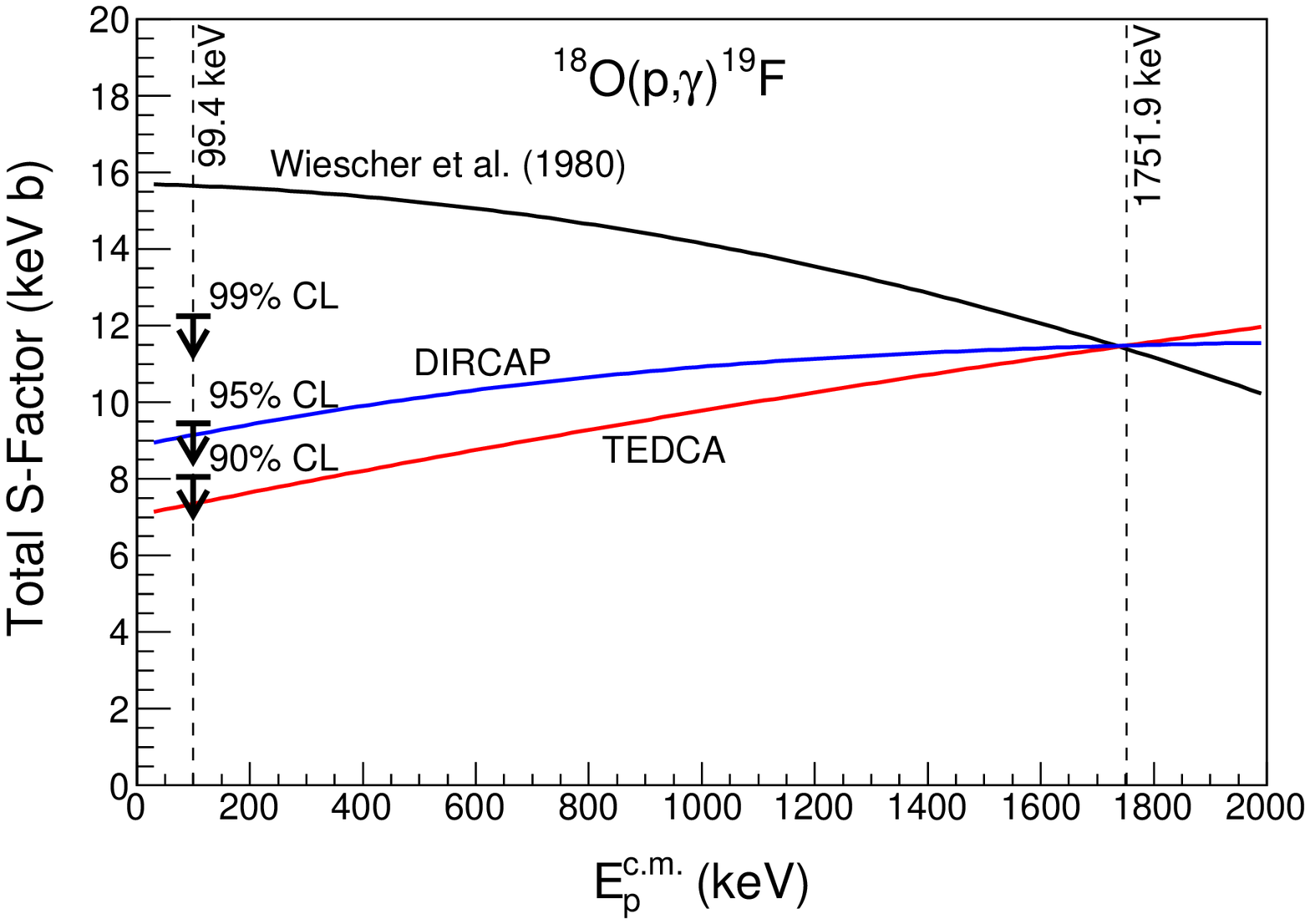}
\caption{\label{fig:totals}(Color online) Total direct capture S-factor for $^{18}$O($p$,$\gamma$)$^{19}$F.  The solid lines represent direct capture model calculations: (black)~\citet{WIE80}; (blue) using code $\mathtt{DIRCAP}$; (red) using code $\mathtt{TEDCA}$; the latter two results are normalized to the measured direct capture cross section at  E$^{\mathrm{lab}}_{p}$~=~1850~keV~\cite{WIE80}.  Our measured upper limits (90$\%$, 95$\%$, 99$\%$ confidence levels) at E$^{\mathrm{c.m.}}_{p}$~=~99.4~keV are displayed as three black arrows.}
\end{center}
\end{figure} 
No direct capture transitions were observed in any of the singles or coincidence spectra accumulated at E$^{\mathrm{lab}}_{p}$~=~105~keV.  An experimental upper limit on the total direct capture cross section was obtained from~\cite{ILI07}:
\begin{equation}
\label{eqn:dcyield}
\mathcal{Y}~=~\frac{\mathcal{N}_{20}}{\mathcal{N}_{p}f^{\mathrm{DC}}_{\gamma}}~=~\frac{1}{\epsilon_{\mathrm{eff}}}\int^{\mathrm{E}_{p}^{\mathrm{c.m.}}}_{\mathrm{E}_{p}^{\mathrm{c.m.}}-\Delta \mathrm{E}} \sigma^{\mathrm{DC}}(\mathrm{E})~\mathrm{dE} 
\end{equation}
where $\mathcal{Y}$ is the measured yield upper limit, $\mathcal{N}_{20}$ is the intensity upper limit of the 2~$\rightarrow$~0 transition from the Bayesian treatment discussed in Sec.~\ref{ss:res}, and $\sigma^{\mathrm{DC}}$(E) is the total direct capture cross section.  This expression assumes that the effective stopping power is approximately constant over the target thickness, as was the case in the present experiment.  The cross section can be rewritten in the form~\cite{ILI07}:
\begin{equation}
\label{eqn:sigmas}
\sigma(\mathrm{E})~=~\frac{\mathcal{S}(\mathrm{E})}{\mathrm{E}}e^{-2\pi\eta}
\end{equation}
where $\mathcal{S}$(E) is the astrophysical S-factor, E is the center-of-mass energy, and $e^{-2\pi\eta}$ is the Gamow factor.  By assuming a nearly constant S-factor over the target thickness, Eqs.~(\ref{eqn:dcyield}$-$\ref{eqn:sigmas}) can be integrated numerically to extract $\sigma$(E) or $\mathcal{S}$(E) from the measured yield.  This set of calculations was performed for the same $\gamma\gamma$-coincidence gate used in Sec.~\ref{ss:res}.  For the total experimental astrophysical S-factor, we found an upper limit of $\mathcal{S}^{\mathrm{DC}}_{\mathrm{total}}~\leq$~8.1~keV~b (90$\%$~CL), corresponding to a direct capture cross section upper limit of $\sigma^{\mathrm{DC}}_{\mathrm{total}}~\leq$~1.8~pb (90$\%$~CL).  Note that these values are nearly independent (within 2$\%$) of the direct capture code used to calculate the branching ratios at E$^{\mathrm{lab}}_{p}$~=~105~keV.

Our experimental total S-factor upper limit (90$\%$ CL) at E$^{\mathrm{lab}}_{p}$~=~105~keV is shown in Fig.~\ref{fig:totals}, along with the values corresponding to the 95$\%$ and 99$\%$ confidence levels.  It is interesting to compare our measured upper limit values with direct capture model calculations.  The black solid curve represents the total S-factor reported by~\citet{WIE80}, while the red and blue solid lines were calculated in the present work using the codes $\mathtt{TEDCA}$~\cite{KRA92} and $\mathtt{DIRCAP}$~\cite{ILI04}, respectively.  The latter two were normalized to the previously measured direct capture cross section at E$^{\mathrm{lab}}_{p}$~=~1850~keV~\cite{WIE80}.  At E$^{\mathrm{lab}}_{p}$~=~105~keV, our measured upper limits are smaller than the prediction of~\citet{WIE80} by about a factor of 2.  The $\mathtt{DIRCAP}$ S-factor (blue line) was only marginally consistent with our experimental upper limit (90$\%$ CL) while the  extrapolation derived from the code $\mathtt{TEDCA}$ (red line) fell within the  90$\%$ confidence level.  
\subsection{Reaction Rates}\label{ss:rr}
Thermonuclear reaction rates for $^{18}$O($p$,$\gamma$)$^{19}$F were calculated using the Monte Carlo method of~\citet{LON10b}.  In the Monte Carlo calculation, using the code $\mathtt{RatesMC}$, we adopted the same nuclear physics input as in Ref.~\cite{ILI10c}, except for the E$^{\mathrm{lab}}_{\mathrm{R}}$~=~95~keV resonance strength, the total direct capture S-factor, the Q-value~\cite{AUD11}, and the resonance energies.  

Based on~\citet{ILI04} , we adopted the $\mathtt{TEDCA}$ extrapolation of the total direct capture S-factor, normalized at E$^{\mathrm{lab}}_{p}$~=~1850~keV~\cite{WIE80}.  For bombarding energies below E$^{\mathrm{c.m.}}_{p}$~=~2.0~MeV, our adopted total S-factor can be expanded around $E$~=~0, with the result:
\begin{align}
\label{eqn:sfit}
&\mathcal{S}(\mathrm{E})~\approx~\mathcal{S}(0)~+~\mathcal{S}'(0)\mathrm{E}~+~\frac{1}{2}\mathcal{S}''(0)\mathrm{E}^{2}  \\
\nonumber &=~7.06~+~2.98\times10^{-3}\mathrm{E}~-~2.60\times10^{-7}\mathrm{E}^{2}~(\mathrm{keV~b}),
\end{align}
where E is the center-of-mass energy.  Note that at low energies, our new direct capture S-factor is significantly smaller than the result reported in~\citet{WIE80}.  Also note that the $\mathcal{S'}$(0) coefficient presented in~\citet{WIE80} was reported incorrectly and should in fact be $\mathcal{S'}$(0)~=~$-0.34$$\times$10$^{-3}$ b~\cite{WIE80b}.  This correction is already applied to the black line displayed in Fig.~\ref{fig:totals}. 

In the present work, we reported on an improved upper limit of the E$^{\mathrm{lab}}_{\mathrm{R}}$~=~95~keV resonance strength, $\omega\gamma$~$\leq$~7.8$\times$10$^{-9}$~eV (90$\%$ CL).  For this particular $^{19}$F level, we may also estimate a lower limit for the resonance strength based on the available resonance properties (see Tab.~\ref{table:levparam}).  The ratio of resonance strengths in the ($p$,$\gamma$) and ($p$,$\alpha$) channels, according to Eq.~(\ref{wgpx}), is given by:  
\begin{equation}
\label{eqn:pgparatio}
\frac{\omega\gamma_{p\gamma}}{\omega\gamma_{p\alpha}}~=~\frac{\Gamma_{\gamma}}{\Gamma_{\alpha}}.
\end{equation}
The ($p$,$\alpha$) strength was measured by~\citet{LOR79}, with the result $\omega\gamma_{p\alpha}$~=~(1.6~$\pm$~0.5)$\times$10$^{-7}$~eV.  An upper limit for the total width of $\Gamma$~$<$~3$\times$10$^{3}$~eV was obtained from the slope of the low-energy edge of the thick-target yield curve~\cite{LOR79}, implying an upper limit of $\Gamma_{\alpha}$~$<$~3$\times$10$^{3}$~eV for the $\alpha$-particle partial width.  Finally, a value of $\Gamma_{\gamma}$~=~(6.0~$\pm$~2.5)$\times$10$^{-1}$~eV was reported for the $\gamma$-ray partial width in~\citet{WIE80}.  With these input values and their associated uncertainties, we found, from Eq.~(\ref{eqn:pgparatio}), a lower limit on the ($p$,$\gamma$) strength of $\omega\gamma_{p\gamma}$~$\geq$~1.3$\times$10$^{-11}$~eV.
\begin{figure}[!bp]
\begin{center}
\begin{picture}(70,260)
\put(-78.5,175){\includegraphics[scale=0.21]{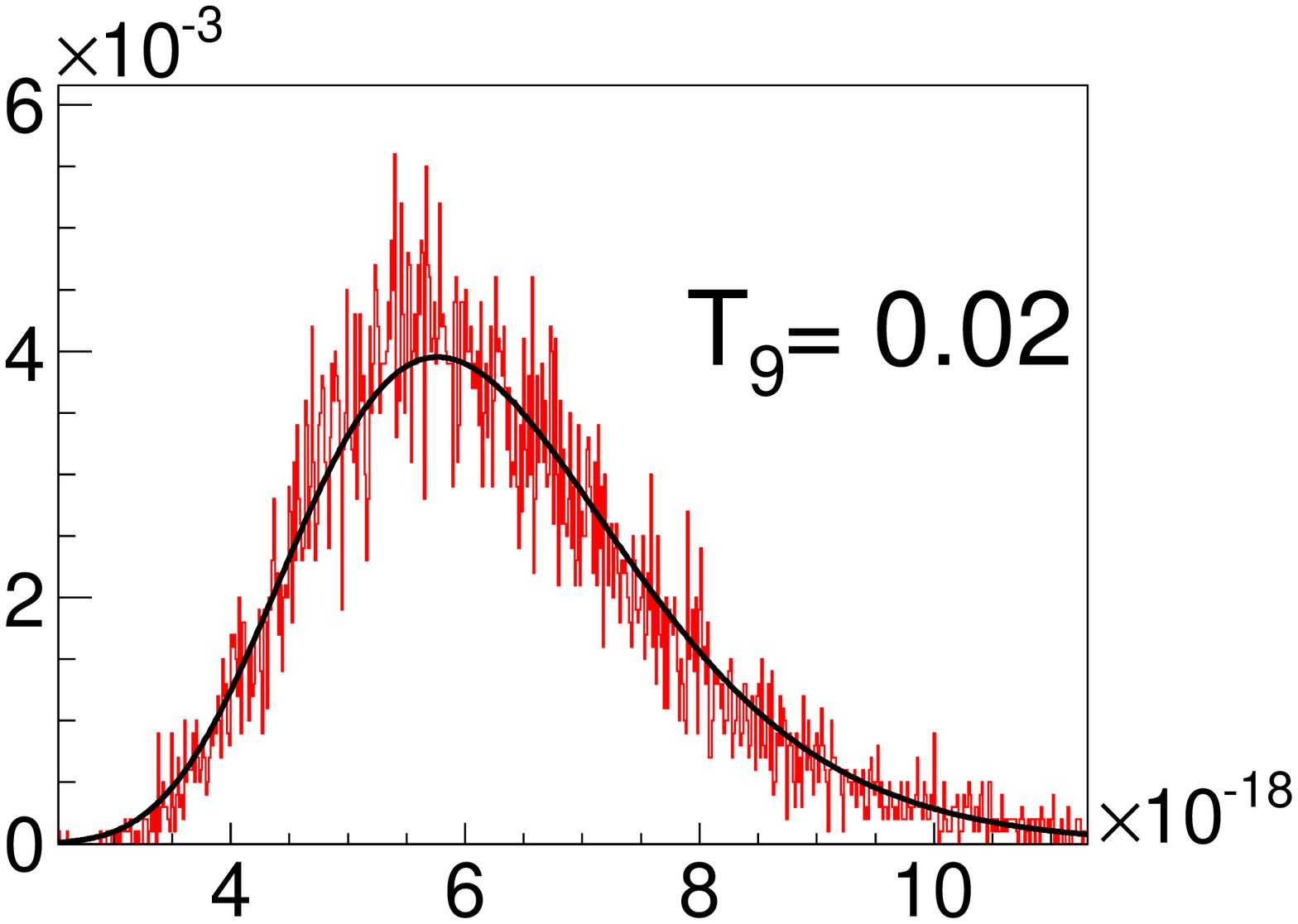}}
\put(5.5,238){(a)}
\put(40.75,175){\includegraphics[scale=0.21]{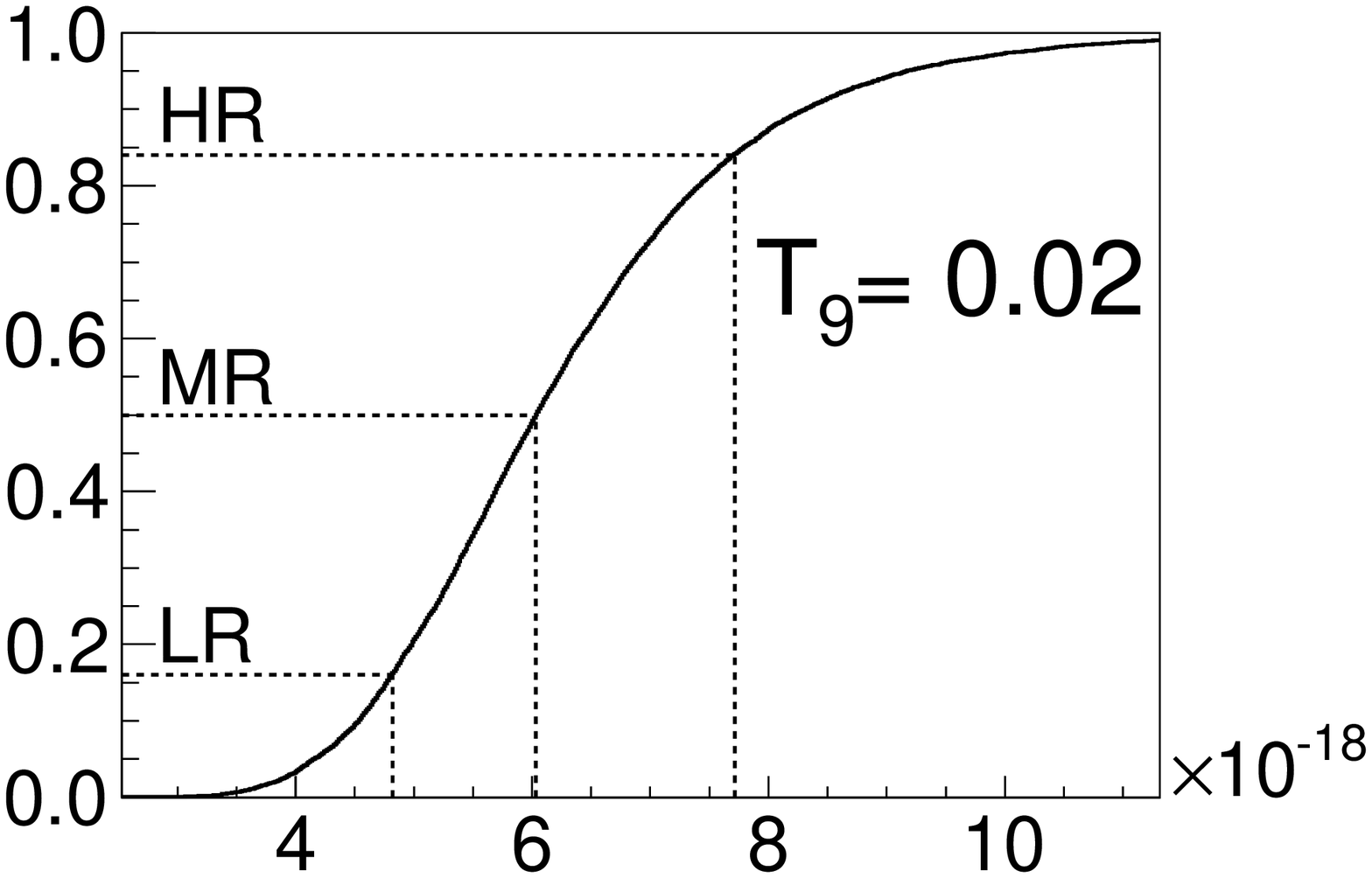}}
\put(124.75,238){(b)}
\put(-93.5,91){\includegraphics[scale=0.21]{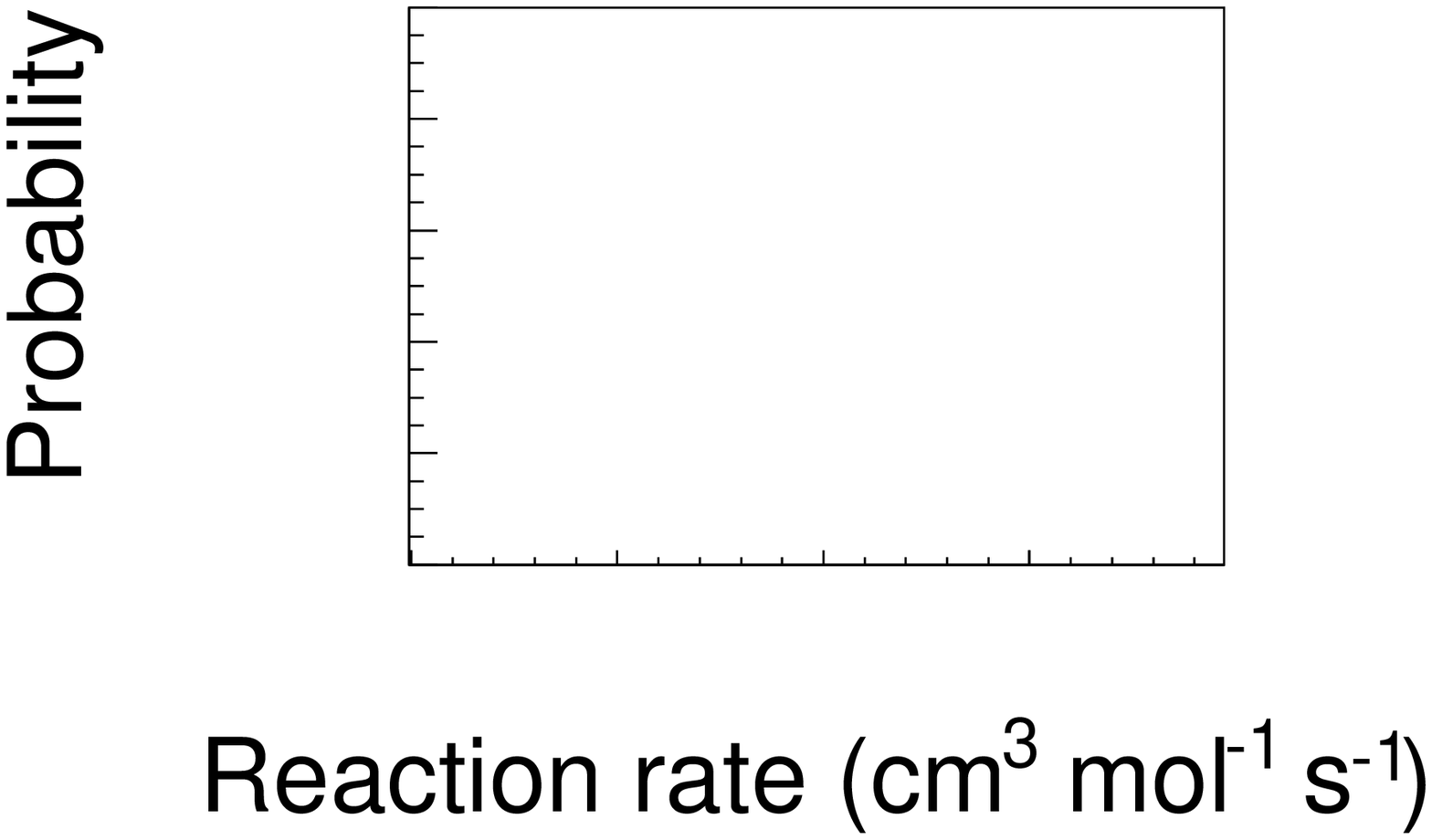}}
\put(-78.5,94.5){\includegraphics[scale=0.21]{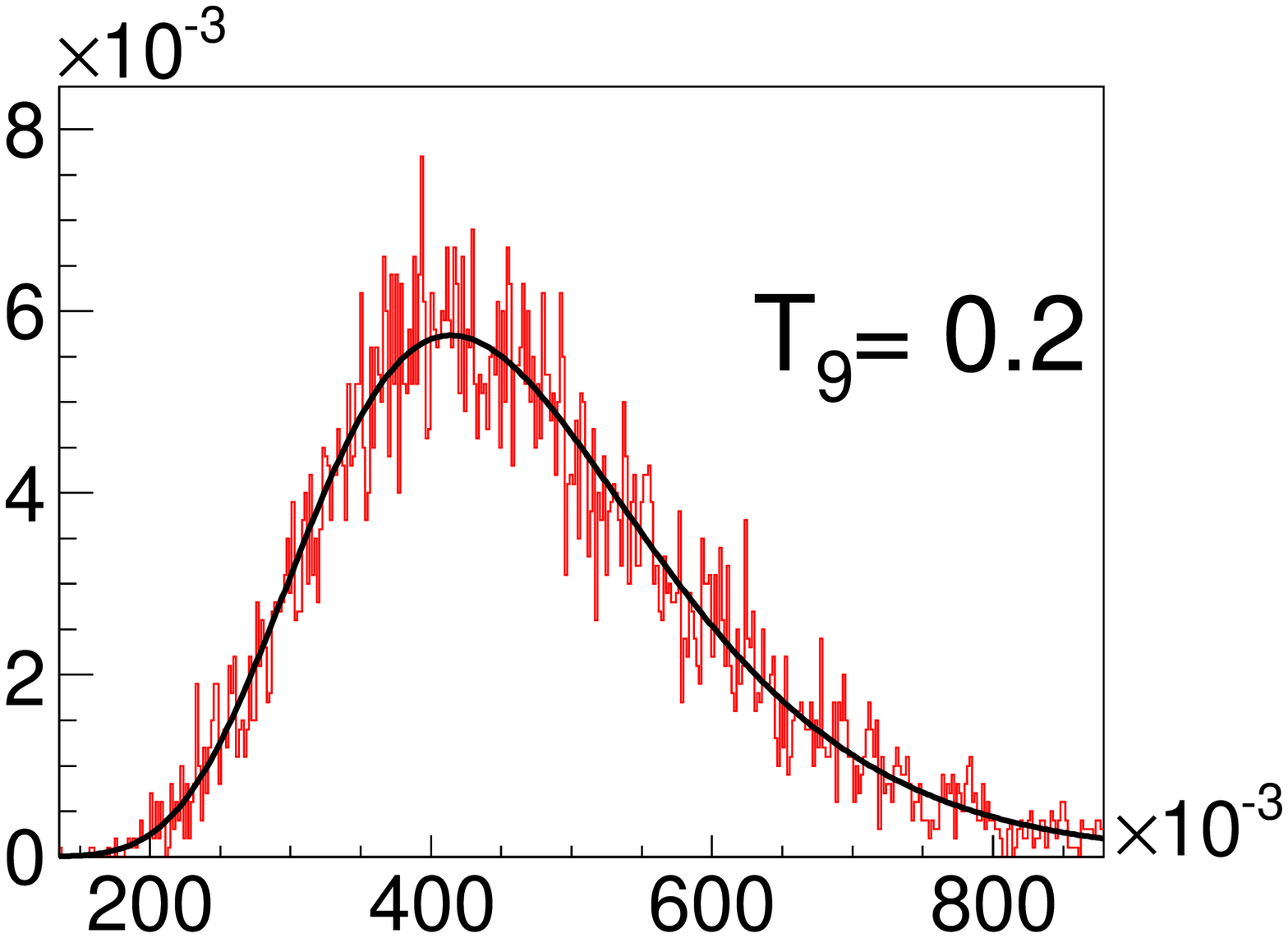}}
\put(5.5,157.5){(c)}
\put(40.75,94.5){\includegraphics[scale=0.21]{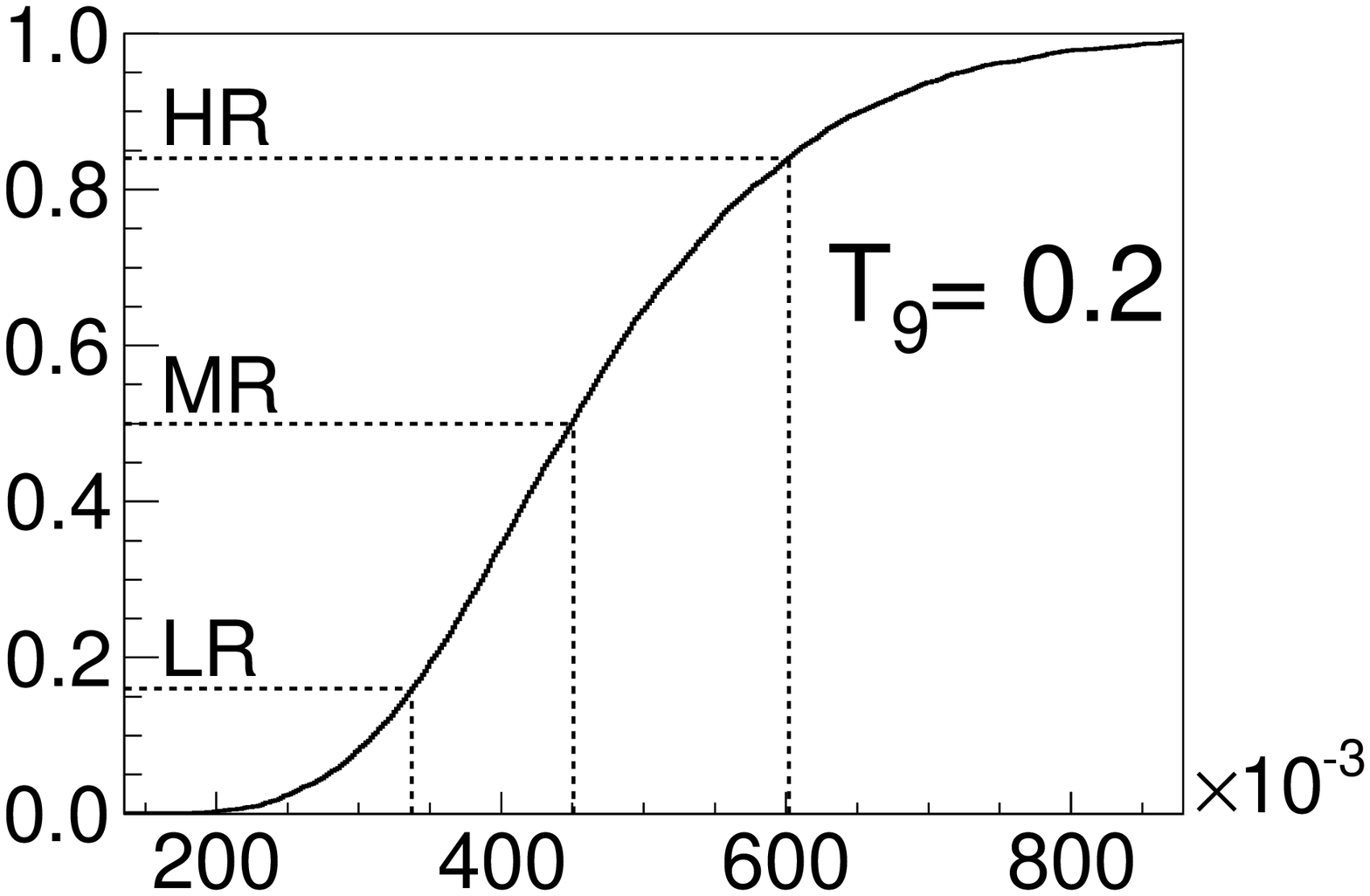}}
\put(124.75,157.5){(d)}
\put(-32,0){\includegraphics[scale=0.21]{RatesMCAxis}}
\put(-78.5,14){\includegraphics[scale=0.21]{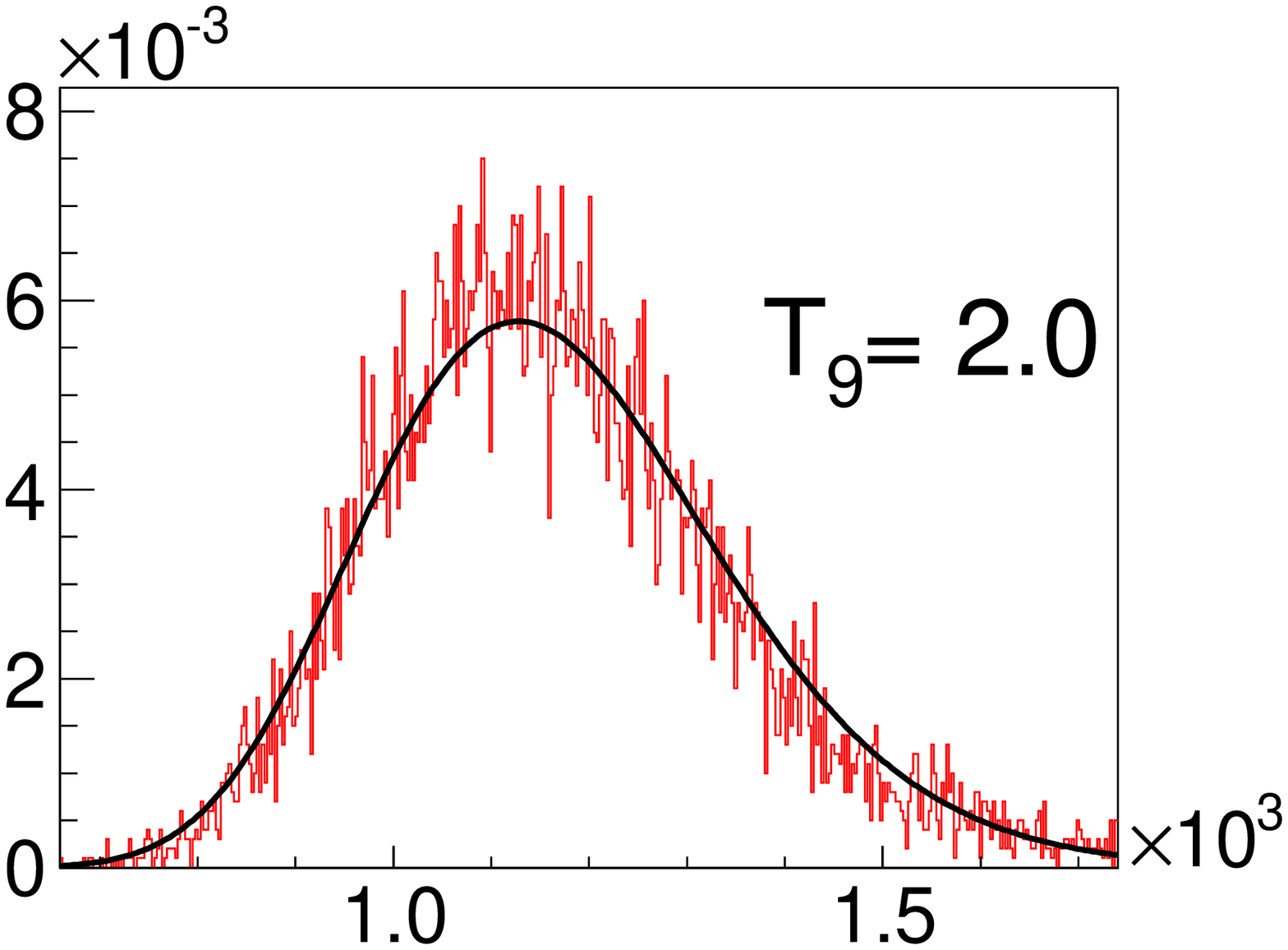}}
\put(5.5,77){(e)}
\put(40.75,14){\includegraphics[scale=0.21]{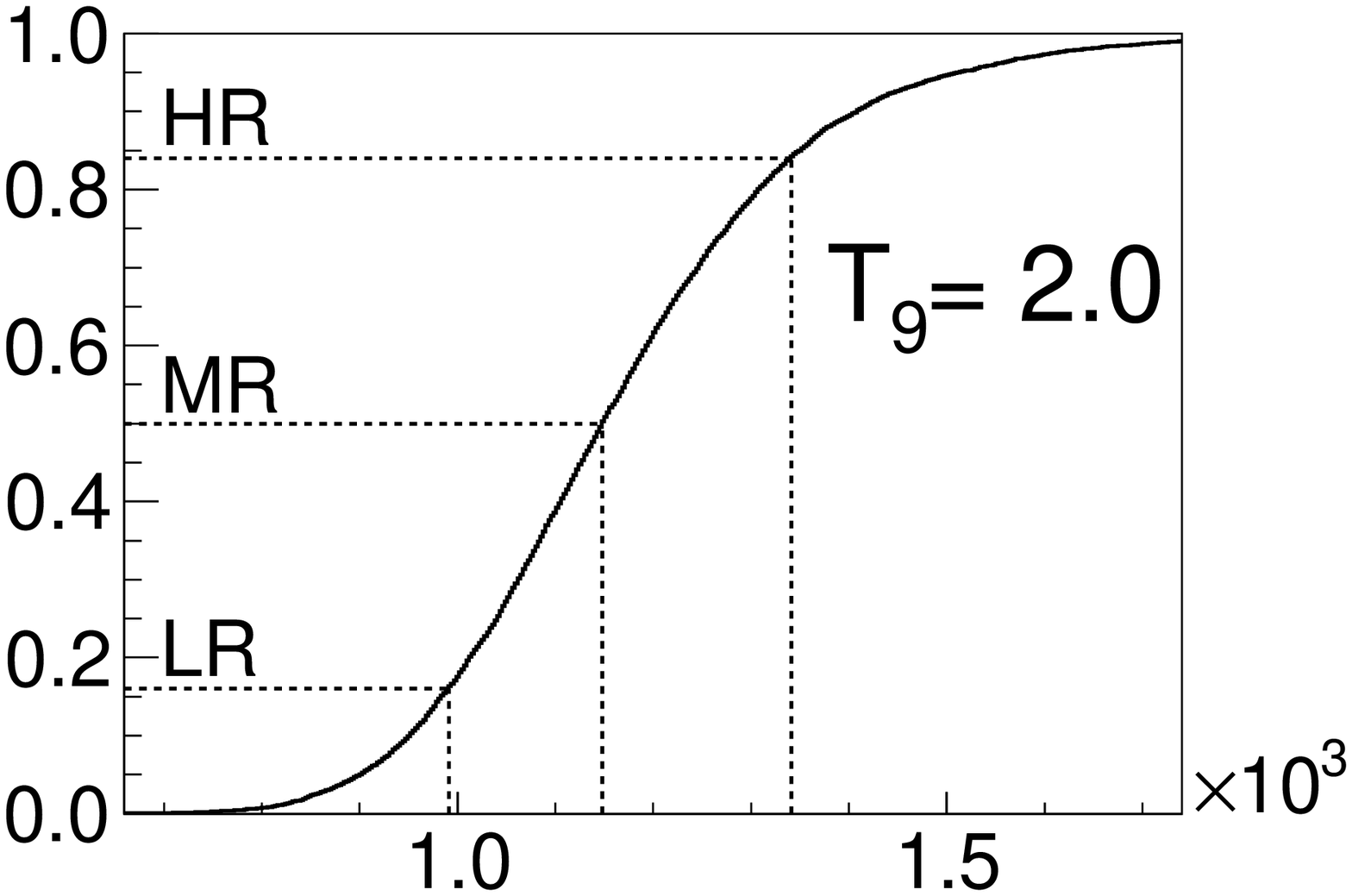}}
\put(124.75,77){(f)}
\end{picture}
\caption{\label{fig:ratesmc}(Color online)(Left) Reaction rate probability density functions (red) for $^{18}$O($p$,$\gamma$)$^{19}$F at 0.02~GK, 0.2~GK, and 2.0~GK generated by the $\mathtt{RatesMC}$ Monte Carlo code.  The lognormal approximations are overlaid in black.  (Right) The corresponding cumulative probability functions used to define the  low, median and high rates as 0.16, 0.5, and 0.84 quantiles, respectively. }
\end{center}
\end{figure}

Since we were able to calculate both an upper and a lower limit on the strength, we estimated a recommended value and a factor uncertainty using the following equations~\cite{LON10b}:
\begin{align}
\label{eqn:expect}
&\omega\gamma = \sqrt{\omega\gamma_{\mathrm{LL}}\times\omega\gamma_{\mathrm{UL}}} = 3.2\times10^{-10}~\mathrm{eV},\\
&\mathrm{f.u.} = \sqrt{\frac{\omega\gamma_{\mathrm{UL}}}{\omega\gamma_{\mathrm{LL}}}} = 25.
\end{align}
In our Monte Carlo procedure, the rate contribution of the E$^{\mathrm{lab}}_{\mathrm{R}}$~=~95~keV resonance strength was found by randomly sampling a lognormal distribution constructed from the mean value and factor uncertainty.  Associating resonance strengths with lognormal distributions is discussed in~\citet{LON10b}. 
\begin{table}[!tp]
\begin{center}
\begin{tabular}{cccccccc} 
\hline\hline
 \multicolumn{2}{c}{T (GK)} & \multicolumn{2}{c}{Low rate} & \multicolumn{2}{c}{Median rate} & \multicolumn{2}{c}{High rate} \\
\hline
\multicolumn{2}{c}{0.010}&\multicolumn{2}{c}{2.923$\times$10$^{-24}$}&\multicolumn{2}{c}{4.967$\times$10$^{-24}$}&\multicolumn{2}{c}{8.785$\times$10$^{-24}$}\\
\multicolumn{2}{c}{0.011}&\multicolumn{2}{c}{2.581$\times$10$^{-23}$}&\multicolumn{2}{c}{4.222$\times$10$^{-23}$}&\multicolumn{2}{c}{7.344$\times$10$^{-23}$}\\
\multicolumn{2}{c}{0.012}&\multicolumn{2}{c}{1.719$\times$10$^{-22}$}&\multicolumn{2}{c}{2.676$\times$10$^{-22}$}&\multicolumn{2}{c}{4.468$\times$10$^{-22}$}\\
\multicolumn{2}{c}{0.013}&\multicolumn{2}{c}{9.299$\times$10$^{-22}$}&\multicolumn{2}{c}{1.370$\times$10$^{-21}$}&\multicolumn{2}{c}{2.181$\times$10$^{-21}$}\\
\multicolumn{2}{c}{0.014}&\multicolumn{2}{c}{4.223$\times$10$^{-21}$}&\multicolumn{2}{c}{5.973$\times$10$^{-21}$}&\multicolumn{2}{c}{8.985$\times$10$^{-21}$}\\
\multicolumn{2}{c}{0.015}&\multicolumn{2}{c}{1.709$\times$10$^{-20}$}&\multicolumn{2}{c}{2.321$\times$10$^{-20}$}&\multicolumn{2}{c}{3.309$\times$10$^{-20}$}\\
\multicolumn{2}{c}{0.016}&\multicolumn{2}{c}{6.180$\times$10$^{-20}$}&\multicolumn{2}{c}{8.151$\times$10$^{-20}$}&\multicolumn{2}{c}{1.113$\times$10$^{-19}$}\\
\multicolumn{2}{c}{0.018}&\multicolumn{2}{c}{6.283$\times$10$^{-19}$}&\multicolumn{2}{c}{8.027$\times$10$^{-19}$}&\multicolumn{2}{c}{1.041$\times$10$^{-18}$}\\
\multicolumn{2}{c}{0.020}&\multicolumn{2}{c}{4.820$\times$10$^{-18}$}&\multicolumn{2}{c}{6.045$\times$10$^{-18}$}&\multicolumn{2}{c}{7.721$\times$10$^{-18}$}\\
\multicolumn{2}{c}{0.025}&\multicolumn{2}{c}{3.235$\times$10$^{-16}$}&\multicolumn{2}{c}{4.041$\times$10$^{-16}$}&\multicolumn{2}{c}{5.168$\times$10$^{-16}$}\\
\multicolumn{2}{c}{0.030}&\multicolumn{2}{c}{8.728$\times$10$^{-15}$}&\multicolumn{2}{c}{1.093$\times$10$^{-14}$}&\multicolumn{2}{c}{1.402$\times$10$^{-14}$}\\
\multicolumn{2}{c}{0.040}&\multicolumn{2}{c}{1.206$\times$10$^{-12}$}&\multicolumn{2}{c}{1.555$\times$10$^{-12}$}&\multicolumn{2}{c}{2.340$\times$10$^{-12}$}\\
\multicolumn{2}{c}{0.050}&\multicolumn{2}{c}{9.349$\times$10$^{-11}$}&\multicolumn{2}{c}{1.227$\times$10$^{-10}$}&\multicolumn{2}{c}{2.056$\times$10$^{-10}$}\\
\multicolumn{2}{c}{0.060}&\multicolumn{2}{c}{9.553$\times$10$^{-9}$}&\multicolumn{2}{c}{1.274$\times$10$^{-8}$}&\multicolumn{2}{c}{1.818$\times$10$^{-8}$}\\
\multicolumn{2}{c}{0.070}&\multicolumn{2}{c}{3.573$\times$10$^{-7}$}&\multicolumn{2}{c}{4.764$\times$10$^{-7}$}&\multicolumn{2}{c}{6.503$\times$10$^{-7}$}\\
\multicolumn{2}{c}{0.080}&\multicolumn{2}{c}{5.495$\times$10$^{-6}$}&\multicolumn{2}{c}{7.304$\times$10$^{-6}$}&\multicolumn{2}{c}{9.891$\times$10$^{-6}$}\\
\multicolumn{2}{c}{0.090}&\multicolumn{2}{c}{4.553$\times$10$^{-5}$}&\multicolumn{2}{c}{6.049$\times$10$^{-5}$}&\multicolumn{2}{c}{8.137$\times$10$^{-5}$}\\
\multicolumn{2}{c}{0.100}&\multicolumn{2}{c}{2.437$\times$10$^{-4}$}&\multicolumn{2}{c}{3.239$\times$10$^{-4}$}&\multicolumn{2}{c}{4.347$\times$10$^{-4}$}\\
\multicolumn{2}{c}{0.110}&\multicolumn{2}{c}{9.505$\times$10$^{-4}$}&\multicolumn{2}{c}{1.263$\times$10$^{-3}$}&\multicolumn{2}{c}{1.692$\times$10$^{-3}$}\\
\multicolumn{2}{c}{0.120}&\multicolumn{2}{c}{2.921$\times$10$^{-3}$}&\multicolumn{2}{c}{3.880$\times$10$^{-3}$}&\multicolumn{2}{c}{5.191$\times$10$^{-3}$}\\
\multicolumn{2}{c}{0.130}&\multicolumn{2}{c}{7.484$\times$10$^{-3}$}&\multicolumn{2}{c}{9.946$\times$10$^{-3}$}&\multicolumn{2}{c}{1.330$\times$10$^{-2}$}\\
\multicolumn{2}{c}{0.140}&\multicolumn{2}{c}{1.663$\times$10$^{-2}$}&\multicolumn{2}{c}{2.211$\times$10$^{-2}$}&\multicolumn{2}{c}{2.957$\times$10$^{-2}$}\\
\multicolumn{2}{c}{0.150}&\multicolumn{2}{c}{3.300$\times$10$^{-2}$}&\multicolumn{2}{c}{4.386$\times$10$^{-2}$}&\multicolumn{2}{c}{5.867$\times$10$^{-2}$}\\
\multicolumn{2}{c}{0.160}&\multicolumn{2}{c}{5.970$\times$10$^{-2}$}&\multicolumn{2}{c}{7.942$\times$10$^{-2}$}&\multicolumn{2}{c}{1.062$\times$10$^{-1}$}\\
\multicolumn{2}{c}{0.180}&\multicolumn{2}{c}{1.581$\times$10$^{-1}$}&\multicolumn{2}{c}{2.102$\times$10$^{-1}$}&\multicolumn{2}{c}{2.814$\times$10$^{-1}$}\\
\multicolumn{2}{c}{0.200}&\multicolumn{2}{c}{3.388$\times$10$^{-1}$}&\multicolumn{2}{c}{4.507$\times$10$^{-1}$}&\multicolumn{2}{c}{6.033$\times$10$^{-1}$}\\
\multicolumn{2}{c}{0.250}&\multicolumn{2}{c}{1.274$\times$10$^{0}$}&\multicolumn{2}{c}{1.694$\times$10$^{0}$}&\multicolumn{2}{c}{2.266$\times$10$^{0}$}\\
\multicolumn{2}{c}{0.300}&\multicolumn{2}{c}{2.932$\times$10$^{0}$}&\multicolumn{2}{c}{3.903$\times$10$^{0}$}&\multicolumn{2}{c}{5.212$\times$10$^{0}$}\\
\multicolumn{2}{c}{0.350}&\multicolumn{2}{c}{5.153$\times$10$^{0}$}&\multicolumn{2}{c}{6.853$\times$10$^{0}$}&\multicolumn{2}{c}{9.134$\times$10$^{0}$}\\
\multicolumn{2}{c}{0.400}&\multicolumn{2}{c}{7.695$\times$10$^{0}$}&\multicolumn{2}{c}{1.020$\times$10$^{1}$}&\multicolumn{2}{c}{1.360$\times$10$^{1}$}\\
\multicolumn{2}{c}{0.450}&\multicolumn{2}{c}{1.037$\times$10$^{1}$}&\multicolumn{2}{c}{1.370$\times$10$^{1}$}&\multicolumn{2}{c}{1.819$\times$10$^{1}$}\\
\multicolumn{2}{c}{0.500}&\multicolumn{2}{c}{1.303$\times$10$^{1}$}&\multicolumn{2}{c}{1.715$\times$10$^{1}$}&\multicolumn{2}{c}{2.269$\times$10$^{1}$}\\
\multicolumn{2}{c}{0.600}&\multicolumn{2}{c}{1.841$\times$10$^{1}$}&\multicolumn{2}{c}{2.387$\times$10$^{1}$}&\multicolumn{2}{c}{3.123$\times$10$^{1}$}\\
\multicolumn{2}{c}{0.700}&\multicolumn{2}{c}{2.464$\times$10$^{1}$}&\multicolumn{2}{c}{3.121$\times$10$^{1}$}&\multicolumn{2}{c}{3.988$\times$10$^{1}$}\\
\multicolumn{2}{c}{0.800}&\multicolumn{2}{c}{3.356$\times$10$^{1}$}&\multicolumn{2}{c}{4.137$\times$10$^{1}$}&\multicolumn{2}{c}{5.129$\times$10$^{1}$}\\
\multicolumn{2}{c}{0.900}&\multicolumn{2}{c}{4.759$\times$10$^{1}$}&\multicolumn{2}{c}{5.709$\times$10$^{1}$}&\multicolumn{2}{c}{6.938$\times$10$^{1}$}\\
\multicolumn{2}{c}{1.000}&\multicolumn{2}{c}{6.916$\times$10$^{1}$}&\multicolumn{2}{c}{8.167$\times$10$^{1}$}&\multicolumn{2}{c}{9.819$\times$10$^{1}$}\\
\multicolumn{2}{c}{1.250}&\multicolumn{2}{c}{1.719$\times$10$^{2}$}&\multicolumn{2}{c}{2.000$\times$10$^{2}$}&\multicolumn{2}{c}{2.380$\times$10$^{2}$}\\
\multicolumn{2}{c}{1.500}&\multicolumn{2}{c}{3.630$\times$10$^{2}$}&\multicolumn{2}{c}{4.213$\times$10$^{2}$}&\multicolumn{2}{c}{4.975$\times$10$^{2}$}\\
\multicolumn{2}{c}{1.750}&\multicolumn{2}{c}{6.403$\times$10$^{2}$}&\multicolumn{2}{c}{7.430$\times$10$^{2}$}&\multicolumn{2}{c}{8.726$\times$10$^{2}$}\\
\multicolumn{2}{c}{2.000}&\multicolumn{2}{c}{9.921$\times$10$^{2}$}&\multicolumn{2}{c}{1.149$\times$10$^{3}$}&\multicolumn{2}{c}{1.342$\times$10$^{3}$}\\
\multicolumn{2}{c}{2.500}&\multicolumn{2}{c}{1.842$\times$10$^{3}$}&\multicolumn{2}{c}{2.129$\times$10$^{3}$}&\multicolumn{2}{c}{2.487$\times$10$^{3}$}\\
\multicolumn{2}{c}{3.000}&\multicolumn{2}{c}{2.798$\times$10$^{3}$}&\multicolumn{2}{c}{3.230$\times$10$^{3}$}&\multicolumn{2}{c}{3.769$\times$10$^{3}$}\\
\multicolumn{2}{c}{3.500}&\multicolumn{2}{c}{3.777$\times$10$^{3}$}&\multicolumn{2}{c}{4.369$\times$10$^{3}$}&\multicolumn{2}{c}{5.130$\times$10$^{3}$}\\
\multicolumn{2}{c}{4.000}&\multicolumn{2}{c}{4.758$\times$10$^{3}$}&\multicolumn{2}{c}{5.507$\times$10$^{3}$}&\multicolumn{2}{c}{6.507$\times$10$^{3}$}\\
\multicolumn{2}{c}{5.000}&\multicolumn{2}{c}{6.600$\times$10$^{3}$}&\multicolumn{2}{c}{7.729$\times$10$^{3}$}&\multicolumn{2}{c}{9.353$\times$10$^{3}$}\\
\multicolumn{2}{c}{6.000}&\multicolumn{2}{c}{(8.727$\times$10$^{3}$)}&\multicolumn{2}{c}{(1.056$\times$10$^{4}$)}&\multicolumn{2}{c}{(1.277$\times$10$^{4}$)}\\
\multicolumn{2}{c}{7.000}&\multicolumn{2}{c}{(1.167$\times$10$^{4}$)}&\multicolumn{2}{c}{(1.411$\times$10$^{4}$)}&\multicolumn{2}{c}{(1.707$\times$10$^{4}$)}\\
\multicolumn{2}{c}{8.000}&\multicolumn{2}{c}{(1.452$\times$10$^{4}$)}&\multicolumn{2}{c}{(1.757$\times$10$^{4}$)}&\multicolumn{2}{c}{(2.125$\times$10$^{4}$)}\\
\multicolumn{2}{c}{9.000}&\multicolumn{2}{c}{(1.718$\times$10$^{4}$)}&\multicolumn{2}{c}{(2.078$\times$10$^{4}$)}&\multicolumn{2}{c}{(2.514$\times$10$^{4}$)}\\
\multicolumn{2}{c}{10.000}&\multicolumn{2}{c}{(2.032$\times$10$^{4}$)}&\multicolumn{2}{c}{(2.458$\times$10$^{4}$)}&\multicolumn{2}{c}{(2.974$\times$10$^{4}$)}\\
\hline\hline
\end{tabular}
\caption{\label{table:reactionrate}Experimental Monte Carlo-based $^{18}$O($p$,$\gamma$)$^{19}$F reaction rates (in units of cm$^{3}~$mol$^{-1}$~s$^{-1}$).  For T~$\geq$~5.5~GK, rates were matched to Hauser-Feshbach results~\cite{GOR08}.}
\end{center}
\end{table}

Our new low, median, and high $^{18}$O($p$,$\gamma$)$^{19}$F reaction rates (corresponding to 0.16, 0.50 and 0.84 quantiles, respectively, of the cumulative reaction rate distribution) are tabulated in Tab.~\ref{table:reactionrate} over a stellar plasma temperature range of 0.01$-$10.00~GK.  Reaction rate probability density functions at a few sample temperatures (0.02, 0.2, 2.0~GK) are displayed as red histograms in Fig.~\ref{fig:ratesmc} (left panel), with the lognormal approximations  shown as black solid lines.  On the right, the corresponding cumulative probability functions are shown with the dashed lines indicating the low, median, and high rates.  It can be seen that a lognormal function approximates the actual Monte Carlo distribution rather well.   

\begin{figure}[!bp]
\begin{center}
\includegraphics[scale=0.45]{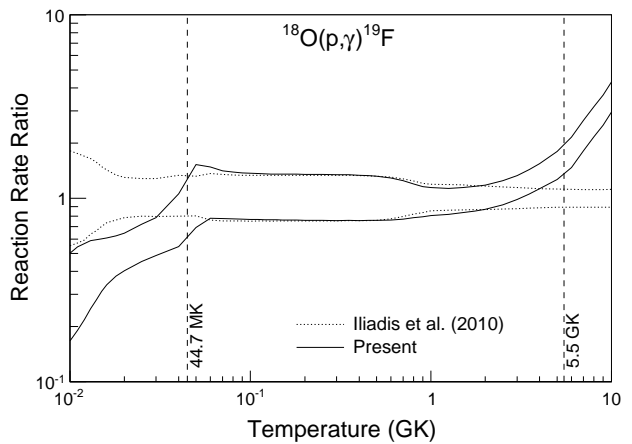}
\caption{\label{fig:buc12vili10pg}Present (solid lines) and previous~\cite{ILI10c} (dotted lines) high and low reaction rates, normalized to the recommended previous rates.  The vertical dashed line at 44.7~MK represents the highest lower limit on CBP temperatures within a low-mass AGB star according to Ref.~\cite{NOL03}.  The vertical dashed line at 5.5~GK represents the temperature at which the experimental rates need to be extrapolated with the aid of Hauser-Feshbach results~\cite{GOR08}.}
\end{center}
\end{figure}

Figure~\ref{fig:buc12vili10pg} compares our new reaction rate with the one published by~\citet{ILI10c}.  The new (solid lines) and previous (dotted lines) high and low rates are normalized to the previous recommended rate~\cite{ILI10c}.  Note that the previous rates contained two small mistakes: (i) an erroneous assignment of J$^{\pi}$~=~(3/2)$^{-}$, and (ii) the incorrectly reported value of $\mathcal{S'}$(0)~=~0.34$\times$10$^{-3}$~b from Ref.~\cite{WIE80}. The dashed vertical line at 44.7~MK indicates the highest temperature threshold at which, according to~\citet{NOL03}, CBP can occur.  The vertical dashed line at 5.5~GK represents the stellar temperature beyond which the rates must be found with the aid of Hauser-Feshbach calculations.  This threshold was computed based on the methodology outlined by~\citet{NEW08}.  

\begin{figure}[!bp]
\begin{center}
\includegraphics[scale=0.45]{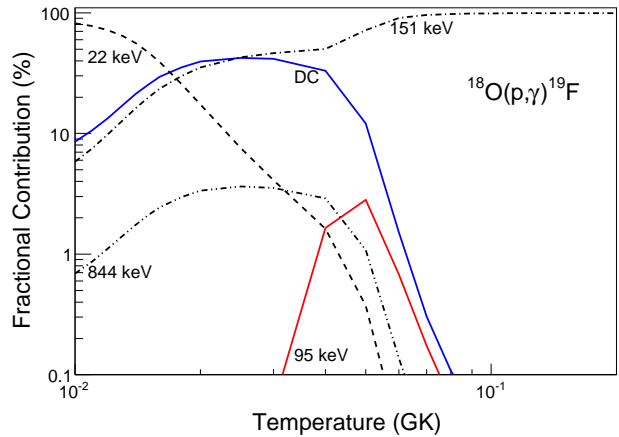}
\caption{\label{fig:fraction}(Color online) Fractional contributions at low temperatures (T$_{9}$~$<$~0.2) to the $^{18}$O($p$,$\gamma$)$^{19}$F reaction rate are shown.  The blue line represents the direct capture contribution, the red line is the E$^{\mathrm{lab}}_{\mathrm{R}}$~=~95~keV contribution, and the dashed, dashed-dotted and dashed-double-dotted lines are the contributions from the E$^{\mathrm{lab}}_{\mathrm{R}}$~=~22~keV, E$^{\mathrm{lab}}_{\mathrm{R}}$~=~151~keV and E$^{\mathrm{lab}}_{\mathrm{R}}$~=~844~keV resonances, respectively.} 
\end{center}
\end{figure} 
The difference, in Fig.~\ref{fig:buc12vili10pg}, between new and previous rates at temperatures below 50~MK can be explained by our lower estimates both for the contributions from direct capture and the resonance at E$^{\mathrm{lab}}_{\mathrm{R}}$~=~95~keV.  Since the new rates are smaller at CBP threshold temperatures compared to the previous results, it is even less likely that the $^{18}$O($p$,$\gamma$)$^{19}$F reaction contributes significantly to the depletion of $^{18}$O observed in stellar atmospheres and presolar grain samples.  The slight increase in the rate above 44.7~MK is dependent upon the calculated E$^{\mathrm{lab}}_{\mathrm{R}}$~=~95~keV strength upper limit and cannot account for observed $^{18}$O depletions.  The difference at temperatures in excess of 5~GK is solely caused by our treatment of the direct capture contribution: the S-factor expansion was artificially cut off at E$^{\mathrm{c.m.}}_{p}$~=~1.0~MeV in previous work~\cite{WIE80,ILI10c}, while in the present work the S-factor is calculated up to energies of E$^{\mathrm{c.m.}}_{p}$~=~2.0~MeV (Fig.~\ref{fig:totals}), resulting in a much higher cutoff value and a significantly increased direct capture contribution.
\begin{figure}[!tp]
\begin{center}
\includegraphics[scale=0.45]{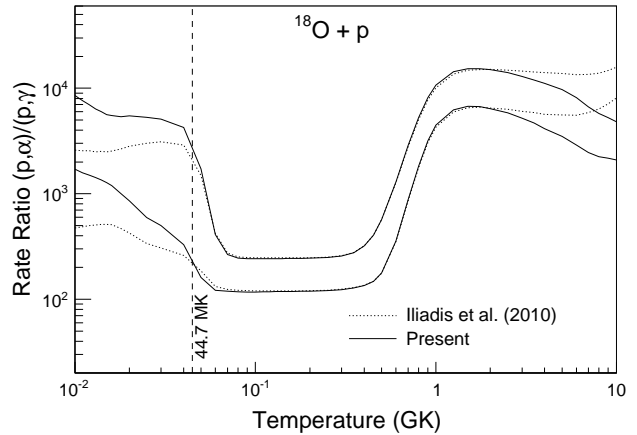}
\caption{\label{fig:buc12vili10pa}Ratios between ($p$,$\alpha$) low and high reaction rates from~\citet{ILI10c} and the present ($p$,$\gamma$) high and low reaction rates, respectively (solid black lines).  The corresponding ratios based solely on the previous rates~\cite{ILI10c} are shown as dotted lines.  The vertical dashed line at 44.7~MK indicates the highest CBP temperature threshold according to~\citet{NOL03}.}
\end{center}
\end{figure} 

The fractional contributions to the total $^{18}$O($p$,$\gamma$)$^{19}$F reaction rates are shown in Fig.~\ref{fig:fraction} for low temperatures.  It can be seen that our lower estimate for the direct capture process (blue solid line) contributes significantly ($>$10$\%$) at temperatures of 0.011$-$0.05~GK.  Our considerably lower estimate for the E$^{\mathrm{lab}}_{\mathrm{R}}$~=~95~keV resonance (red solid line) yields a negligible contribution ($<$3$\%$) in the temperature region relevant to cool bottom processing.  The dashed lines represent the reaction rate contributions of the other resonances (E$^{\mathrm{lab}}_{\mathrm{R}}$~=~22, 151 and 844~keV) that influence the rate at stellar plasma temperatures below 0.2~GK. 

The ratio of $^{18}$O($p$,$\alpha$)$^{15}$N and $^{18}$O($p$,$\gamma$)$^{19}$F high and low rates is shown in Fig.~\ref{fig:buc12vili10pa}.  The dotted lines are based on the results of Ref.~\cite{ILI10c} alone, while the solid lines incorporate the new $^{18}$O($p$,$\gamma$)$^{19}$F rates.  For the temperature region relevant to CBP, the established ($p$,$\alpha$) rate~\cite{ILI10c} exceeds the ($p$,$\gamma$) rate by a factor of 5100$-$1700 over the temperature range 0.03$-$0.05~GK.  From our improved E$^{\mathrm{lab}}_{\mathrm{R}}$~=~95~keV resonance strength upper limit and our refined direct capture S-factor, we support the conclusion that the ($p$,$\gamma$) reaction does not contribute significantly to the overall $^{18}$O destruction at temperatures suggested for CBP to occur in low-mass AGB stars.  Future efforts to study $^{18}$O depletion by CBP in AGB stars should focus on direct measurement of the $^{18}$O($p$,$\alpha$)$^{15}$N reaction at low energies.
\section{Conclusion}\label{ss:conc}
A study of the $^{18}$O($p$,$\gamma$)$^{19}$F reaction was performed at the Laboratory for Experimental Nuclear Astrophysics (LENA).  A new resonance strength upper limit of $\omega\gamma~\leq$~7.8$\times$10$^{-9}$~eV (90$\%$ CL) for the E$^{\mathrm{lab}}_{\mathrm{R}}$~=~95~keV resonance was measured that improves upon the previous ($p$,$\gamma$) upper limit published by~\citet{VOG90} by about half an order of magnitude.  Our data also allow for a significant improvement of the total direct capture S-factor prediction.  Our direct capture S-factor amounts to about half of the previously accepted value at low energies~\cite{WIE80}.  With this experimental information, new Monte Carlo-based reaction rates for $^{18}$O($p$,$\gamma$)$^{19}$F are derived.  We find that the new reaction rates in the hypothesized CBP temperature regime are even smaller than previously assumed.  Clearly, $^{18}$O depletion in low-mass AGB stellar atmospheres and some presolar oxide grains is dominated by the competing $^{18}$O($p$,$\alpha$)$^{15}$N reaction.  Future studies of $^{18}$O depletion by cool bottom processing in low-mass AGB stars should focus on direct measurement of the ($p$,$\alpha$) reaction at low energies. 
\acknowledgments
The authors would like to thank TUNL technical staff members J. Addison, B. P. Carlin, J. Dunham,  B. Jelinek, P. Mulkey, R. O'Quinn, and C. Westerfeldt.  Special thanks to A. L. Sallaska, L. N. Downen and B. M. Oginni.  Additional thanks to R. Longland, J. R. Newton, and C. W. Arnold.  The authors would also like to thank A. Coc (CSNSM, Orsay) and G. Angelou (Monash University).  This work was supported in part by the US Department of Energy under Contract no. DE-FG02-97ER41041. Additional support was provided for MQB by the DOE NNSA Stewardship Science Graduate Fellowship under Grant no. DE-FC52-08NA28752.
\end{document}